# Quantum-mechanical effects in photoluminescence from thin crystalline gold films

A. R. Bowman[1], A. Rodríguez Echarri[2,3], F. Kiani[1], F. Iyikanat[2], T. V. Tsoulos[1], J. D. Cox[4,5], R. Sundararaman[6,7], F. Javier García de Abajo[2,8], and G. Tagliabue[1]*

1. Laboratory of Nanoscience for Energy Technologies (LNET), STI, École Polytechnique Fédérale de Lausanne (EPFL), Lausanne 1015, Switzerland
2. ICFO–Institut de Ciencies Fotoniques, The Barcelona Institute of Science and Technology, 08860 Castelldefels (Barcelona), Spain
3. MBI–Max-Born-Institut, 12489 Berlin, Germany
4. POLIMA–Center for Polariton-driven Light–Matter Interactions, University of Southern Denmark, Campusvej 55, DK-5230 Odense M, Denmark
5. Danish Institute for Advanced Study, University of Southern Denmark, Campusvej 55, DK-5230 Odense M, Denmark
6. Department of Materials Science & Engineering, Rensselaer Polytechnic Institute, 110 8th Street, Troy, New York 12180, USA
7. Department of Physics, Applied Physics, and Astronomy, Rensselaer Polytechnic Institute, 110 8th Street, Troy, New York 12180, USA
8. ICREA–Institució Catalana de Recerca i Estudis Avançats, Passeig Lluís Companys 23, 08010 Barcelona, Spain
*corresponding author: giulia.tagliabue@epfl.ch

**Abstract**

Luminescence constitutes a unique source of insight into hot carrier processes in metals, including those in plasmonic nanostructures used for sensing and energy applications. However, being weak in nature, metal luminescence remains poorly understood, its microscopic origin strongly debated, and its potential for unravelling nanoscale carrier dynamics largely unexploited. Here, we reveal quantum-mechanical effects emanating in the luminescence from thin monocrystalline gold flakes. Specifically, we present experimental evidence, supported by first-principles simulations, to demonstrate its photoluminescence origin when exciting in the interband regime. Our model allows us to identify changes to the measured gold luminescence due to quantum-mechanical effects as the gold film thickness is reduced. Excitingly, such effects are observable in the luminescence signal from flakes up to 40 nm in thickness, associated with the out-of-plane discreteness of the electronic band structure near the Fermi level. We qualitatively reproduce the observations with first-principles modelling, thus establishing a unified description of luminescence in gold and enabling its widespread application as a probe of carrier dynamics and light-matter interactions in this material. Our study paves the way for future explorations of hot-carriers and charge-transfer dynamics in a multitude of material systems.

**Introduction**

Luminescence from semiconductors following steady-state photoexcitation has been known since ancient times [1]. Today, semiconductor luminescence is widely employed as a non-invasive probe of diverse phenomena, including the dating of rocks, the assessment of solar cell efficiencies and monitoring of chemical reactions [2–5]. These applications are possible because the processes that cause luminescence are well understood. In contrast, luminescence from metals was first observed in only 1969 [6], due to the signal being orders of magnitude weaker than in most semiconductors. Photon emission from metals has recently received increased attention in the context of plasmonic nanostructures, which promise to revolutionise industries including healthcare, sensing and energy [7–9] due to the ability of plasmon-generated hot carriers to dramatically elevate local electronic temperatures, augment weak luminescence processes from molecules and increase solar cell absorption [10,11]. Steady-state luminescence has a unique potential to shed light on hot carrier processes in plasmonic systems. Although it is more accessible from an experimental viewpoint, this process has received less attention than two-photon photoluminescence with pulsed lasers [12–15]. Nevertheless, steady-state luminescence from metals has been employed for fundamental nanoscale studies [16,17], monitoring surface and electronic temperature [18–21], probing gold-molecule interactions [22,23], and monitoring charge transfer [24,25]. Despite its usefulness, uncertainty remains around the origin of emitted light, particularly whether it is due to inelastic light scattering or recombination of electrons and holes (the latter termed photoluminescence, PL), with many theoretical and experimental studies debating these possibilities over the last 50 years [14,17,26–33] (see also a recent summary by Cai et al. [34]). This debate is further complicated by the Purcell enhancement of emission at specific wavelengths of light that resonates with plasmonic modes of the metal structure (which dominates the predicted spectrum in several works [17,35–40]), the position of the excitation wavelength relative to the interband transition threshold and spatial confinement. To the best of our knowledge, a full understanding of steady-state luminescence from metals following interband excitation without the participation of such resonant excitations is still lacking, thus hindering its applications as an effective probe.

Here, we study photon emission from 13 nm to 113 nm thick monocrystalline, atomically flat, gold flakes with the (111) surface exposed [41]. These samples allow us to probe the relationship between photon emission and nanoscale confinement without surface roughness or plasmonic enhancement, meaning that the conclusions we draw can be generally applied to any metal, not only those operating in the plasmonic regime. Our study reveals that, when illuminating in the interband regime, the long-wavelength photon emission is independent of the excitation wavelength: conclusive evidence that this signal is due to photoluminescence rather than inelastic scattering. In addition, we demonstrate that gold luminescence can be used as a probe of local temperature using only the Stokes signal (i.e. signal at longer wavelengths than excitation wavelength) when exciting at 488 nm wavelength. To further understand the emission, we employ photon re-absorption to reveal that there is minimal charge diffusion after photoexcitation prior to photon emission. This enables us to present a model of luminescence that includes photon re-absorption and first-principles calculations parameterised by density-functional theory (DFT), producing results in good agreement with PL experiments. Gold PL (when exciting in the interband regime) is shown to consist of two key components, both resulting from the recombination of excited d-band holes with unexcited electrons: pre-scattered luminescence close to the excitation energy, and longer wavelength post-scattered luminescence. Our model of bulk luminescence allows us to identify that as the flake thickness is reduced below 40 nm, quantum mechanical confinement effects on states near the Fermi level cause an increase in pre-scattered luminescence at longer wavelengths (when compared to thick flakes), which we justify via first-principles modelling. Finally, we explore luminescence signals when exciting in the intraband regime. By invoking scaling arguments, we propose that intraband luminescence is in fact due to inelastic scattering. Our results provide a comprehensive theory of gold photoluminescence that is readily applicable to other metals, realise a more accessible form of nanoscale thermometry, and resolve a 50-year-old paradox on the origin of luminescence in gold.

**Main**

We synthetized bare monocrystalline gold flakes of 113 nm down to 13 nm thicknesses on quartz following Kiani et al. [41] (with (111) surface exposed), with lateral dimensions greater than 5 μm in all samples. We studied 22 flakes and present representative results here. In all measurements, we used focused laser beams near the sample centre (i.e., far from edges which could cause plasmonic enhancement [17]). A schematic of the measurement and a white-light image of a gold flake also showing a focused laser spot on it are presented in Figures 1a and 1b, respectively. To confirm that luminescence signals did not originate from defects or edges, we excited several points across the sample surface and found identical results. Additionally, we carried out spatially resolved measurements, which confirmed signals originated solely from the region excited by the laser for all excitation wavelengths employed (see Supplemental Note 1). We also verified that the luminescence observed from our flakes was identical to monocrystalline gold fabricated by an entirely different synthesis method [42] (Supplemental Note 2). Therefore, we are confident that our measurements present photon emission from gold itself without plasmonic enhancement effects being present, in contrast to most studies that specifically focus on the role of surface plasmons [43]. We also note that signals from flakes were unchanged over several months, demonstrating that our samples were stable.

We present photon emission from an 88 nm gold flake as a function of excitation wavelength in the interband (488 nm and 532 nm) and intraband (785 nm) regimes in Figure 1c (signal normalized by the absorbed laser power). Broadband emission over the entire visible region is observed when exciting in the interband regime consistent with previous studies [6]. Interestingly, for the 785 nm laser excitation, the measured signal is weaker, but on the same order of magnitude, refuting previous reports that emission in this region is purely due to plasmonic effects [29,31]. We note the peak at 900 nm observed with 785 nm excitation is due to an experimental artefact – it is also observed on a silver mirror. When exciting in the interband regime (488 nm and 532 nm), we find that the emission spectra per absorbed photon overlap perfectly at long wavelengths, (i.e., $\lambda_{em} \gtrsim 600$ nm, Figure 1c). This leads us to two important conclusions. Firstly, because the shape of the long wavelength emission spectrum is independent of the excitation wavelength, carriers must undergo relaxation prior to radiative recombination. Thus, an inelastic scattering mechanism (Figure 1d, i.) can be ruled out in this regime. In support of this conclusion, we further note that signals do not overlap when plotted as a function of the energy shift relative to the incident laser energy (Supplemental Note 3). We can therefore conclude purely experimentally that photon emission at long wavelengths must be due to PL when exciting in the interband regime (Figure 1d, ii.). This is in agreement with studies demonstrating that the lifetime of PL following interband excitation is on the order of 50 fs (while inelastic scattering is instantaneous) [44,45]. Secondly, as the absolute signals overlap in the long wavelength region, we can infer that carriers contributing to PL follow the same decay pathway, independent of excitation wavelength. In contrast, photon emission close to the excitation wavelength does not overlap when exciting with different lasers in the interband regime, consistent with previous observations of PL from gold nanoparticles [17].

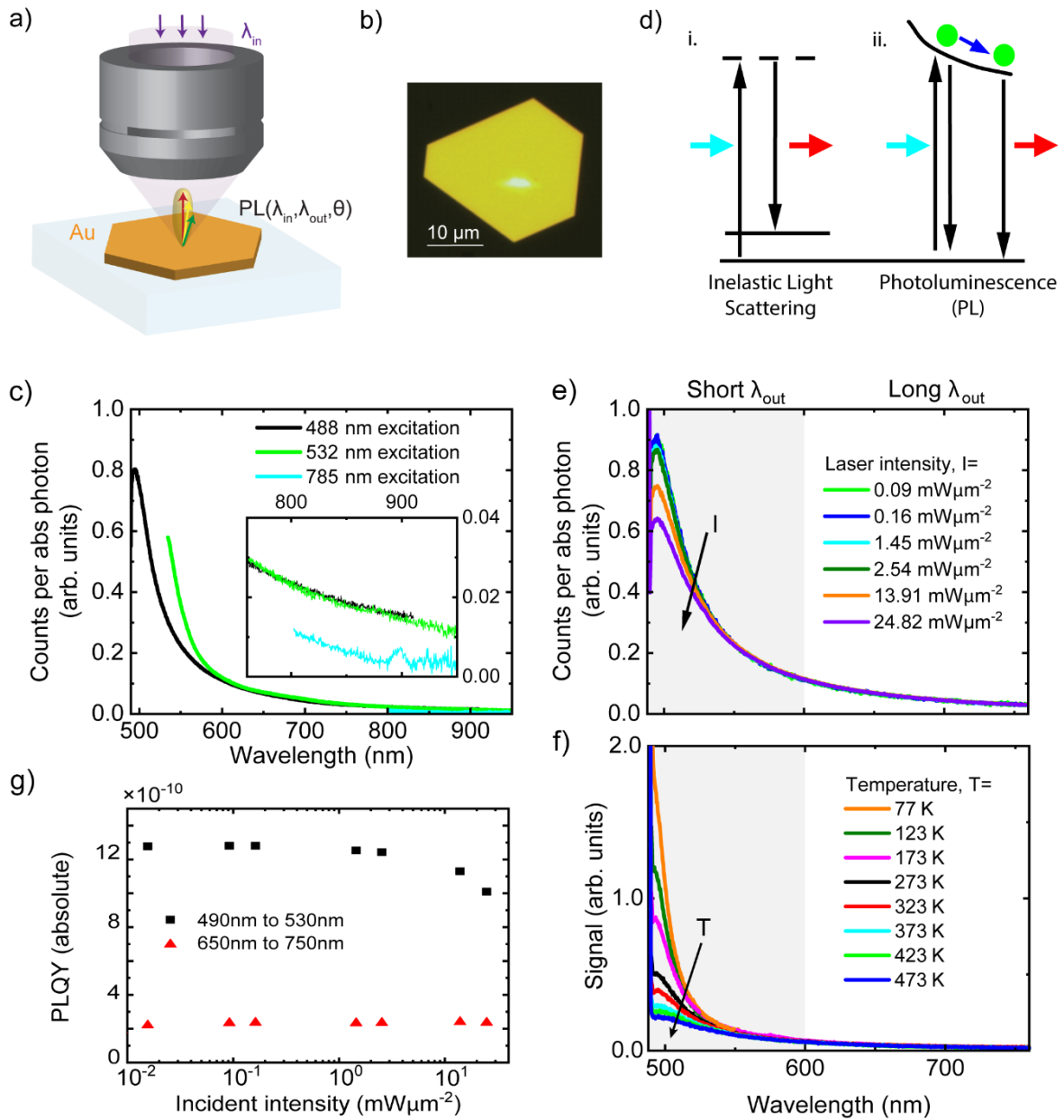

*Figure 1. a) Schematic of the measurement process including the incident laser spot and emitted light. b) White light reflection image of a gold flake in a microscope (79 nm thickness), with a 488 nm wavelength laser incident on the central portion of the flake. c) Photoluminescence of an 88 nm thick gold flake normalized per absorbed photon as a function of excitation wavelength. The inset expands the long wavelength region. Incident intensities are 1.449 mWµm$^{-2}$/3.334 mWµm$^{-2}$/0.063 mWµm$^{-2}$ for 488 nm/532 nm/785 nm laser excitation wavelength. d) Schematics of i. inelastic light scattering and ii. photoluminescence processes. Photoluminescence can occur both before and after scattering of an excited electron (green dots), while inelastic light scattering is mediated by a virtual state (dashed line),*

*without any relaxation process taking place prior to photon emission. e) and f) spectral dependence of the photoluminescence of a 110 nm thick gold flake for e) different laser excitation powers at room temperature, and f) sample temperature for a constant excitation power of 0.072 mWμm$^{-2}$. g) External photoluminescence quantum yield (in absolute scale) as a function of excitation power for a 110 nm thick flake. In e)-g) the excitation wavelength is 488 nm and the grey shaded area corresponds to the defined "short $\lambda_{em}$" range.*

We varied the incident laser power on the sample to measure how the signal changed. For 488 nm excitation, we found the short wavelength emission ($\lambda_{\text{out}} < 600$ nm) PL reduced in magnitude (per absorbed photon) for higher incident laser powers (Figure 1e). Importantly, we do not observe a similar effect for 532 nm or 785 nm excitation (for comparable rates of absorbed photons). This change is fully reversible upon reducing the 488 nm laser power. We explain this behaviour by noting that the same effect is observed upon sample heating (with fixed, low laser power), as is shown in Figure 1f. By comparing the signal with laser irradiation and with external heating, we can measure the local sample temperature non-invasively for different laser powers. We find that the lattice temperature can increase by ~200 K for thin samples (<20 nm) for the laser intensities used, while the lattice temperature only increases by 70 K for thick samples (>50 nm), as shown in Supplemental Note 4.

Considering the distinct behaviour of the PL signal at short and long wavelengths for 488 nm excitation (with/without temperature dependence), we estimate the photoluminescence quantum yield (PLQY) separately for these two parts (Figure 1g). Our external PLQY estimates of $\sim 10^{-10}$ are in good agreement with Mooradian's original estimates and well below those of plasmon-enhanced nanoparticle emission [6,46]. As reported before, apart from temperature effects, the PLQY remains constant when varying the excitation power [17], so we are operating in the linear response regime. Importantly, this rules out nonlinear processes (for example, involving two excited carriers or the response of an already excited system, simplifying modelling [30]) and also implies that the signal is due to the recombination of an excited electron with an unexcited hole (or vice-versa). See Supplemental Note 5 and our previous work for further discussion [47].

We modelled the luminescence from the gold using a formalism of dipoles emitting throughout the material [48]. As is schematically depicted in Figure 2a, the measured PL is a combination of the dipole emission strength (based on the material's band structure), charge transport prior to luminescence, and gold's absorption coefficients. Regarding charge transport, we consider the two limiting cases of: i) neglecting transport i.e. excitation and emission take place at the same location; and ii) a maximally delocalized model where equal emission occurs from every point inside the film. Within the far-field limit, and assuming that the luminescence spectrum is position independent, these models can be written as

$$PL_{\text{external}} = PL_{\text{internal}} \left(\frac{I_{\text{in}}}{\hbar\omega_{\text{in}}}\right) \int_0^d dz_0 \, f_{\text{abs}}(z_0, \lambda_{\text{in}}) \, f_{\text{emit}}(z_0, \lambda_{\text{out}}) \qquad (1a)$$

for local PL and

$$PL_{\text{external}} = PL_{\text{internal}}(\omega_{in}, \omega_{out}) \left(\frac{I_{in}}{\hbar\omega_{\text{in}}}\right) \int_0^d dz_0' \, f_{\text{abs}}(z_0', \lambda_{\text{in}}) \int_0^d dz_0 \, f_{\text{emit}}(z_0, \lambda_{\text{out}}) \qquad (1b)$$

for maximally delocalized PL. Here, $PL_{\text{external}}$ is the measured (external) luminescence per unit area and unit wavelength, $PL_{\text{internal}}$ is the probability of photon emission at wavelength $\lambda_{\text{out}}$ per photon absorbed at incident wavelength $\lambda_{\text{in}}$ and unit of emission wavelength, $\frac{I_{\text{in}}}{\hbar\omega_{\text{in}}}$ is the rate of photons incident on the sample per unit area (expressed as the ratio of the laser intensity to the incident photon energy), $f_{\text{abs}}(z_0, \lambda_{\text{in}})$ the fraction of incident photons absorbed per unit length across the film (noting that $\int f_{\text{abs}}(z_0, \lambda_{\text{in}}) \, dz_0$ gives the total photon absorption in the material) and $f_{\text{emit}}(z_0, \lambda_{\text{out}})$ describes the probability that a photon emitted at position $z_0$ inside the metal escapes the material (encapsulating photon re-absorption processes) and is detected. We present a rigorous derivation of all terms in equation 1, parameterised by the refractive index of gold [49], in Supplemental Note 6. Incidentally, the internal PL has been defined here to connect with the terminology used by the semiconductor luminescence community, though we note that care should be taken when attributing a physical meaning to this quantity, which is ultimately proportional to the dipole emission strength.

To model the luminescence according to equation 1, we need to (i) quantify the sample absorption and (ii) decide whether the local or delocalized model correctly describes the situation in gold. We performed reflection/transmission measurement to obtain the absorption spectra which are in agreement with theoretical calculations (see Supplemental Note 7 for absorption model and experimental results). As recently reported [50], a way to spatially resolve the photon emission distribution is to measure the ratio of the PL spectra per absorbed photon when exciting the sample from the same and opposite sides as the signal collection (see schematic in Figure 2b). In the case of localized PL, we should obtain a ratio significantly larger than 1 (that can be modelled using equation 1a). Conversely, for the maximally delocalized case a ratio of 1 is obtained for all metal film thicknesses. We present the measured and calculated ratios for 47 nm and 14 nm flakes in Figure 2c. Strong support is obtained from experiment to our local PL model, demonstrating minimal charge redistribution prior to PL, in agreement with theoretical predictions (which are in turn supported by recent experimental results [51,52]). This analysis also confirms that PL originates from the bulk of the film rather than originating from the surfaces (which predicts significantly higher ratios, for example for the 47 nm thick flake presented ratios greater than 10, not shown on the plot). We find similar results for films of all thicknesses, though we note that for the thinner films (<25 nm) this ratio is close to 1 with or without charge motion because charges are more uniformly excited. This analysis, which we believe is the first measurement revealing the spatial origin of gold luminescence, demonstrates the ability of photon re-absorption to resolve charge distribution at nanometer length scales.

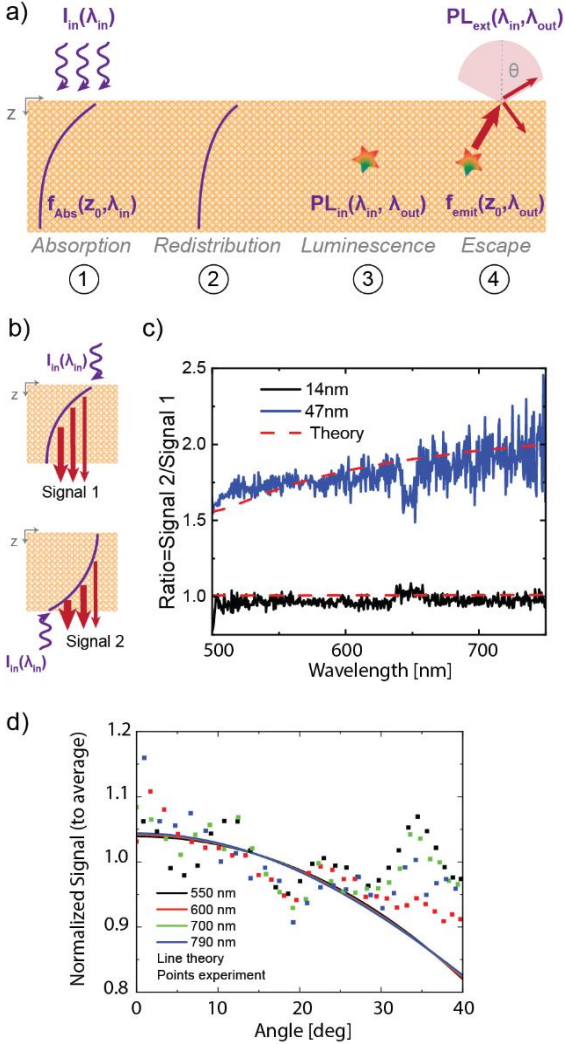

*Figure 2. a) Schematic of the processes that occur following photon absorption (equation 1), starting from point 1 and proceeding to point 4. b) Schematic of measurement approach – a gold flake is excited from above and below, and the photoluminescence signal is recorded from the same bottom side in both measurements (Signal 1 and Signal 2, respectively). c) Wavelength dependence of the ratio between these two signals indicated in b), alongside calculations assuming that photoluminescence emission occurs from the locations where charges are generated. We present results for flakes of 47 nm and 14 nm thickness. For exciting and collecting the same (opposite) side, excitation power is 0.057 mWµm$^{-2}$ (0.052 mWµm$^{-2}$). Signals are normalised to the number of incident photons. d) Angle-resolved luminescence (normalised by average value) at 550 nm, 600 nm, 700 nm and 790 nm (± 10 nm) emission wavelengths, alongside predicted angular dependence, for 488 nm excitation on an 88 nm thick flake at 0.604 mWµm$^{-2}$ excitation intensity.*

To confirm our model of photon emission, we recorded the back-focal-plane (BFP) PL signal as a function of angle. A schematic of the emission configuration is presented in Figure 2a (step 4). We plot the BFP PL signal for 550 nm, 600 nm, 700 nm and 790 nm emission wavelengths from an 88 nm thick flake in Figure 2d. We note this signal was extremely difficult to measure due to competing luminescence from the objective lens (which was much more strongly focused in BFP measurements and did not present a similar issue in real space measurements, see Methods). Despite this, on all plots we observe agreement between the experiment and our prediction based on $f_{\text{emit}}$.

We now consider the role of film thickness both in photon absorption and escape probability. We present the PL signal for gold microflakes with decreasing thicknesses ($d$) from 113 nm down to 13 nm in Figure 3a. Two main changes occur: the short wavelength signal decreases, while the long wavelength signal increases. The sharp peaks at ~ 500 nm for thinner samples are quartz substrate

Raman resonances. To explore whether the signal changes with thickness are due to $f_{\text{abs}}$ and $f_{\text{emit}}$, without having competing contributions from the spectral shape of $PL_{\text{internal}}$, in Figure 3b we present the ratio of the measured luminescence spectra relative to the signal measured for a 113 nm flake. In our model this is equivalent to the ratio:

$$\frac{\int_d f_{\text{abs}} f_{\text{emit}} dz}{\int_{d=113nm} f_{\text{abs}} f_{\text{emit}} dz}. \tag{2}$$

This factor is overlaid as dashed lines on Figures 3b and 3c for thick ($\geq 47$ nm) and thin ($< 47$ nm) flakes, respectively. Good agreement is observed between experiment and theory down to 47 nm. However, for flakes thinner than approximately 40 nm, some deviations are observed: specifically, we record less signal at short wavelengths ($< 550$ nm) and more signal at intermediate wavelengths (550 nm to 700 nm) than our model anticipates. However, theory and experiment still agree at sufficiently long wavelengths. Additionally, resonance features appear for thinner flakes, as can be observed in Figure 3a. We first discuss flakes with thicknesses greater than 40 nm, which follow our theory well, and then explore thin flakes further.

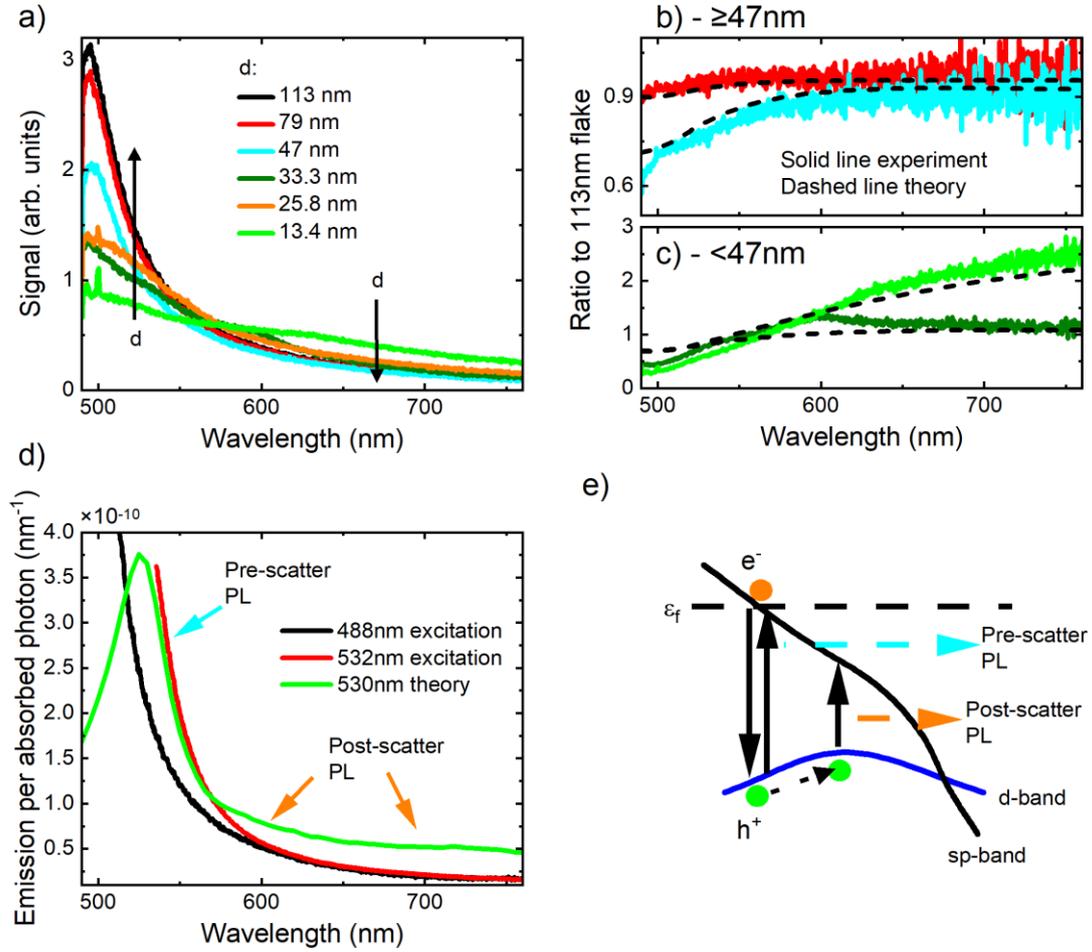

*Figure 3. a) Measured photoluminescence for different flake thicknesses d under 488 nm excitation wavelength. b) and c) ratio of measured signals presented in a) to the signal obtained from the 113 nm flake and separated into b) thick and c) thin flakes respectively. Dashed lines show calculated profiles based on the model presented in the main text (equation 2). Curve colours in b) and c) are coordinated with a). d) Calculated internal luminescence (using an 88 nm thick flake data) for 488 nm and 532 nm light excitation, alongside simulated internal luminescence for 530 nm wavelength excitation. e) Schematic of the processes leading to photoluminescence, where $\varepsilon_f$ indicates the Fermi energy and $h^+/e^-$ represents the excited hole/electron following photoexcitation. The incident laser power on the sample is 0.162 mWμm$^{-2}$/3.334 mWμm$^{-2}$ for 488 nm/532 nm excitation wavelength in all panels.*

We present $PL_{internal}$ for 488 nm and 532 nm excitation in Figure 3d, which we calculate by dividing the measured luminescence by the known factors in equation 1. The internal PL is approximately an order of magnitude stronger than the measured PL (the wavelength resolved external PLQY is on the order of 10$^{-11}$), and has a stronger signal at short wavelengths and weaker signal at long wavelengths when compared to the shape of the measured PL (see Supplemental Information Note 6). To better understand PL from gold, we developed a model parameterised by the output of DFT simulations that have already reproduced the optical constants of gold in the visible regime and finds quantitative agreement with pump-probe measurements on gold nanoparticle solutions [53–56]. Here, we implemented a model of luminescence [57,58] which was first applied to metals by Apell et al. [59] and more recently employed by several groups in experimental studies, notably Cai et al. [17] (see Supplemental Note 8 for model details). Our model advances on the literature by including both direct and phonon-assisted luminescence transitions, momentum conservation and a steady-state excitation regime. We note analytical expressions using an approach equivalent to ours have been derived and applied in the case of intraband excitation [60]. Importantly, our model is fully quantitative, allowing for a direct comparison of both spectra and intensity. We present simulated results for 530 nm excitation wavelength in Figure 4d. We obtain agreement between experiment and theory: luminescence that decreases in intensity further from the excitation wavelength and becomes relatively flat at longer wavelengths. This plot is on an absolute, rather than a relative scale, meaning that we reproduce the ratios of scattering rates to luminescence rates well. Importantly, to successfully reproduce experimental signals, we include two PL components in our modelling: unscattered carriers producing PL close to the excitation wavelength, and scattered carriers producing PL at longer wavelengths. In Supplemental Figures S8 and S10 we show that, when exciting in the interband regime nearly all luminescence originates from d-band holes recombining with unexcited electrons (including emission at wavelengths longer than 700 nm – we note that, unlike absorption, interband PL can still occur in this region [61]). A schematic of the processes governing gold PL is presented in Figure 3e. To the best of our knowledge this is the first time unscattered carrier luminescence has been shown to play a key role in steady-state luminescence from a material. We note a larger discrepancy between our experiment and theory at wavelengths longer than 600 nm, though still well within the error for these experiments and calculations. We discuss this discrepancy further in Supplemental Note 8 section 5, noting it could be due to additional loss processes in the metal not accounted for in simulations, or numerical artefacts. We also simulated luminescence with 488 nm excitation wavelength, finding less good agreement between experiment and theory. Specifically, we found that the DFT model starts to include lower energy d-bands do not play a role in our experiments. This is not unreasonable as the error for a specific band in energy is approximately 0.2 eV for state-of-the-art DFT models (this point is further discussed in Supplemental Note 8).

We now discuss PL from ≤ 40 nm thick flakes. We present the PL signal per absorbed photon for several flake thicknesses for 488 nm and 532 nm excitation wavelengths in Figure 4a. For thick flakes, the signal per absorbed photon with 488 nm and 532 nm excitation are identical at emission wavelengths ($\lambda_{\text{out}}$) larger than approximately 600 nm (Figure 1c and related discussion). However, the point where these signals start to overlap shifts to longer wavelengths as the flake thickness is reduced: for 33.3 nm it appears at 750 nm, while for 13.4 nm this overlap occurs at around 800 nm. Weak resonances are also observed for thin flakes (see Figure 3a). Similar resonances were observed from such flakes in two-photon photoluminescence from thin flakes by Großmann et al. [13]. These resonances are inelastic in nature, as can be seen in Figure 4a: the resonance peak position shifts depending on the excitation wavelength. In fact, resonance peaks are found to have a constant energy shift relative to the energy of the excitation wavelength, as shown in Figure 4b for a 33.3 nm flake. We also observed comparable signals when exciting the film from the air side or through the quartz substrate (Supplemental Note 9), meaning these features can only be explained by changes to the electronic environment in the bulk material.

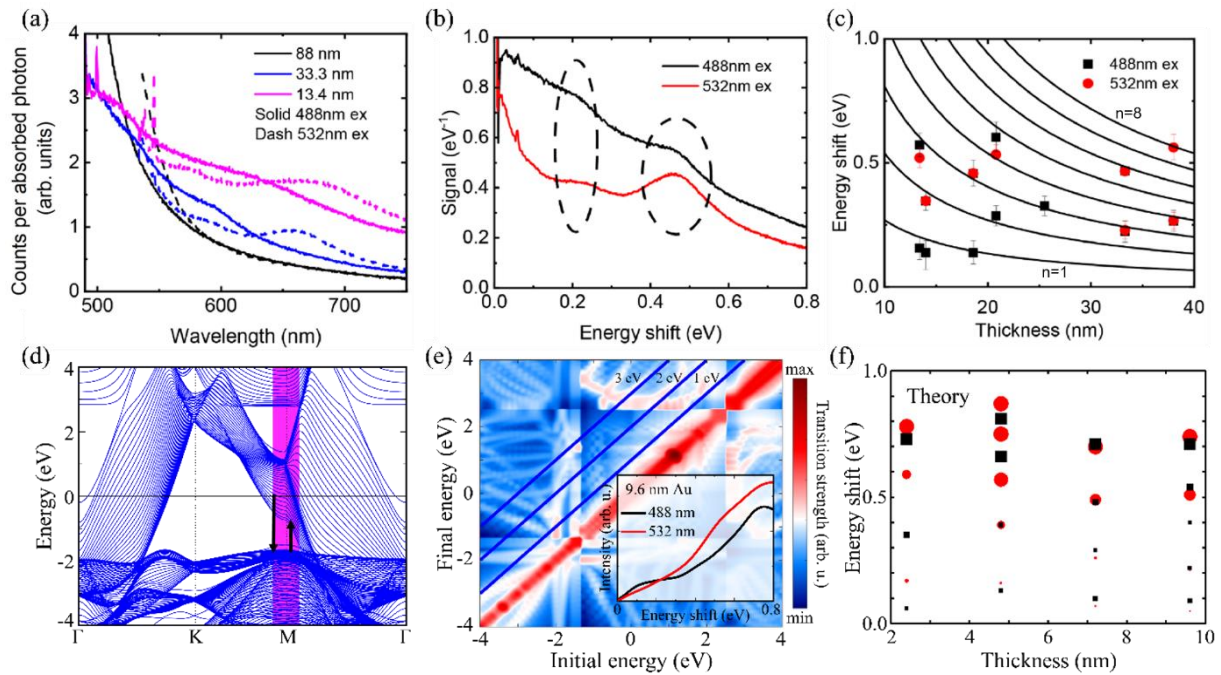

*Figure 4. a) Luminescence signal normalized per absorbed photon for flakes of 88 nm, 33.3 nm and 13.4 nm thicknesses following excitation with 488 nm and 532 nm wavelength light (0.322 mWμm$^{-2}$ and 0.549 mWμm$^{-2}$ excitation intensities respectively). b) Luminescence signal for the 33.3 nm flake, as a function of energy shift for 488 nm (1.449 mWμm$^{-2}$) and 532 nm (3.334 mWμm$^{-2}$) excitation. c) Energy shift of resonance peaks as a function of sample thickness for both 488 nm (black squares) and 532 nm (red circles), with the fitting described in the main text with different black lines corresponding to different n values, as marked on figure. d) Electronic band dispersions of 40 atomic layers of Au (111). The vicinity of the M point of the Brillouin zone is indicated by the pink area and suggested pre-scattered luminescence process overlaid in black arrows, which indicate excitation and subsequent relaxation of a hole. e) Dipole matrix elements of 40 atomic layers of Au (111) averaged over the pink area of the Brillouin zone shown in d) as a function of initial and final state energies. Luminescence intensities estimated using the dipole matrix elements are shown in the inset as a function of energy shift for 488 nm and 532 nm excitation. f) Variation of the predicted intensity (size of the black squares and red circles indicate the strength of the corresponding transitions emerging as features in the spectra of the inset in e) ) and energy shift of the resonance peaks obtained from first-principles calculations as a function of gold film thickness.*

When we record the energy shift of the resonance peak (i.e. $\hbar(\omega_{in} - \omega_{out})$, with $\omega_{out}$ at the resonance peak) for different flake thicknesses (Figure 4c), we find that, for a given thickness, energy shifts appear approximately at multiples of the lowest shift. For example, the energy shifts for a 33.3 nm flake are 0.23 eV and 0.46 eV, suggesting different orders of the same process. A plausible origin of the changes to luminescence lies in unscattered hole PL occurring at longer wavelengths, caused by small changes in the electronic states available close to the Fermi level. Two additional points support this interpretation. Firstly, at sufficiently long wavelengths the signal still tends towards what we expect from our model presented in equation 1 (Figure 3c), implying that the change is not due to a modification in the electron-electron or electron-phonon scattering rates (which are in any case local parameters), but rather to a change in the band structure experienced by excited holes. Secondly, we only observe these effects in ≤ 40 nm thick flakes. While the mean free path of holes in the d-band is only around 2 nm, the mean free path of electronic states close to the Fermi level is around 40 nm [51], indicating that these states are influenced by the finite thickness of the film. We find that different qualitative models can fit the data well (see Supplemental Note 10 for discussion). In Figure 4c, we fit all data points by considering the breakdown of periodic boundary conditions in the vacuum direction (for electronic states close to the Fermi level). This gives energy shifts of $\frac{nhv_f}{2t}$, where $n$ is an integer, $h$ is Planck's constant and $v_f$ is the Fermi velocity (see Supplemental Note 10). We fit the experimental data well with $v_f = 1.3 \times 10^6$ ms$^{-1}$, close to previous DFT values of $1.4 \times 10^6$ ms$^{-1}$ [51].

We further support our claim of this effect being due to the breakdown of periodic boundary conditions with DFT calculations based on the thickness-dependent electronic band structure of Au (111) and the ensuing transition dipole matrix elements [62,63]. Electronic band dispersions of 40 layers of Au (111) are shown in Figure 4d. As expected, the bands near the Fermi level are spread in energy due to the breakdown in periodic boundary conditions. The pink area around the M point of the Brillouin zone is highlighted as this is the relevant region that stands out as a candidate source of the experimentally observed spectral features originating in the piling up of many electronic bands around the Fermi level. We overlay black arrows (describing hole excitation and subsequent de-excitation) indicating the form of pre-scattered luminescence we believe to be responsible for the signals we observe. In Figure 4e, the dipole matrix elements averaged over the area around the M points are shown as a function of the initial and final state energies. Transitions corresponding to 1, 2, and 3 eV energy jumps lie along the diagonal blue lines shown in the figure. Subsequently, using the formula given in Supplemental Note 11, we produce a rough estimate of the contribution of the dipole matrix elements to the luminescence intensity that occurs in the system when the material is illuminated with 488 nm and 532 nm excitation wavelength and illustrated it in the inset of Figure 4e. It is worth noting that in our calculations, we collect all the transitions between the excitation and emission wavelength with equal energy for transitions around the M point. Additionally, we illustrate the resonance peaks in Figure 4f (sizes of symbols show the relative strengths of the resonance amplitudes) as a function of thickness and optical energy shift. We find qualitative agreement between experimental and theoretical results: the energy shifts are at the same order of magnitude and we anticipate the same energy shifts for 488 nm and 532 nm excitation, especially for thicker flakes. Nevertheless, our calculations predict more resonance peaks at multiples of the lowest peak that are not revealed in the experiment, potentially due to the long wavelength signals being too weak for this to be observable in experiment. Although the layer thicknesses that we consider in our calculations are smaller than in experiment, it is clear that resonance peaks organized as a function of energy shift tend to overlap as the thickness increases, as also observed in the measurements.

Finally, we discuss the intraband luminescence (785 nm excitation, Figure 1c). We apply an approach similar to that presented in Figure 3c to explore how the signal changes with sample thickness (Supplemental Note 12). Importantly, when exciting at 785 nm, our model of how the signal changes with thickness agrees well for all samples, and we see no additional resonance features or significant deviations from theory (contrary to Figure 3c for the thinner films). This is consistent with our interpretation of the resonances as pre-scattered photoluminescence: we predict pre-scattered luminescence to be much weaker than post-scattered luminescence for intraband transitions. We also calculate the internal intraband PL signal. Surprisingly, our DFT-parameterised model does not agree well with the experimentally recorded PL. Specifically, the experimental results are approximately a factor of five larger than those predicted by both the DFT-parameterised PL model and by simple scaling arguments, and the spectral profile of the luminescence is also significantly different. Therefore, our data demonstrates that, when exciting in the intraband regime, luminescence cannot be explained by photoluminescence. Instead, here we suggest that it is due to inelastic light scattering. In Supplemental Figure S17d, we estimate the approximate magnitude of this effect based on our experiments, allowing future investigations to calculate its strength relative to PL. Our data suggests that photoluminescence is dominant when exciting in the interband regime, while inelastic light scattering dominates excitation in the intraband regime. This final insight unifies the majority of the literature on luminescence from gold – most studies claiming inelastic light scattering have used 785 nm excitation [29], which corresponds to a photon energy below the interband threshold, while most studies claiming PL have employed interband excitation energies [17].

**Conclusion**

In summary, we present a comprehensive study of photon emission from gold monocrystalline flakes following excitation with continuous wave lasers. We give direct evidence from luminescence and present a first-principles parameterized model to show that this signal is photoluminescence when exciting in the interband regime. Importantly, we show that a significant component of photoluminescence arises from pre-scattered carriers. By comparing this model with different flake thicknesses, we demonstrate that additional quantum mechanical effects become important when flakes are thinner than 40 nm, which we explain by noting that for thin flakes the electronic band structure changes near the Fermi level (due to this thickness being comparable with the mean free path of Fermi-level electrons). Finally, by careful quantitative analysis based on our model, we propose that luminescence when exciting in the intraband regime cannot be explained by photoluminescence, but is rather due to inelastic light scattering. We believe that our study offers a blueprint for studying luminescence in other metals besides gold, provides the first comprehensive model of continuous wave luminescence calculations in metals incorporating insight from density-functional theory, and reveals quantum mechanical effects on the luminescence of metals.

**Experimental Methods**

**Sample Fabrication:** Monocrystalline gold flakes were fabricated following Kiani et al. (using the bottom surface rather than the inter-substrate surface, see reference) [41]. Here, samples were fabricated on quartz substrates (University Wafer) as glass photoluminescence was found to outcompete that from thin gold structures. Substrates were cleaned in ultrasonic baths of ethanol followed by de-ionized water (both for 15 minutes) prior to sample fabrication. Samples were washed in ethanol followed by de-ionized water, and then dried with a nitrogen gun, prior to any measurements.

**Luminescence Measurements:** 488 nm measurements presented in Figures 1d-f, Figure 4a and all 532 nm measurements were carried out on Renishaw inVia Raman Microscope RE04, for 488 nm using a Coherent sapphire 488 SF NX and 532 nm Nd-YAG (RL532100) excitation and a Lecia HC PL FLUOTAR 100x long working distance objective (NA=0.75). All other measurements were carried out on a NanoMicroSpec-Transmission™ (NT&C) microscope, further adapted to enable Raman spectroscopy (which gave excellent agreement with the commercial Renishaw system). Specifically, 488 nm (Matchbox) and 785 nm (Thorlabs L785H1 diode mounted into a Thorlabds LDM56/M, current and temperature controlled by LDC205C and TED200C) continuous-wave lasers were coupled into a Nikon Eclipse Ti2 inverted microscope. Here the objective was a 60x Nikon S Plan Fluor (NA=0.7). The laser beam was passed through laser line bandpass filters (Thorlabs FL488-1 and Semrock LL01-488 for 488 nm excitation/Semrock LL01-785 for 785 nm excitation) prior to impacting on the sample. For both excitation wavelengths two Thorlabs notch filters (NF488-15/NF785-33 for 488 nm and 785 nm excitation) were used to remove any laser from the collected signal. The signal was coupled to a Princeton Instruments Spectra Pro HRS-500 spectrometer and recorded on a PIXIS 256 camera. All signals were radiometrically calibrated using a calibration lamp (Ocean Optics HL-3P-INT-CAL) coupled to an integrating sphere (Thorlabs 2P3/M), see 'photoluminescence quantum yield' section for further details. Spot sizes were assumed diffraction limited on the Renishaw inVia Raman Microscope. For the NT&C system, the laser spot size was recorded using a camera and fitted by a two-dimensional Gaussian. Spot size here is as the intensity that falls to $\frac{1}{e^2}$ of the maximum intensity. When calculating the relative emission intensity of signals with different laser excitation wavelengths (as in Figures 1c and 4a) we first performed measurements using one laser, and then carried out a calibration measurement. Without removing the lamp or the calibration sphere, we then changed the collection path filters for measurements with a second laser, and we then recorded the calibration on this adjusted setup. This means that even if our PLQY estimates contain inaccuracies, the relative intensity measurements of different laser excitation wavelengths is robust.

**Laser Power Determination:** was recorded using a Thorlabs S170C or S130C power meter (using the microscope power meter for any measurements at the sample position).

**Back-focal-plane:** Two lenses were placed between the image plane of the microscope and the spectrometer to enable back-focal-plane measurements, following methods described by Kurvits et al. [64]. Initially, a signal of a larger order of magnitude than the gold signal was observed when no sample was present in the system, which we attribute to the objective lens glowing. To remove this objective signal from BFP measurements we closed an iris in a real space plane between the sample and detector, to exclude signals except those originating from the region surrounding the sample. However, we still observed a small background signal from the objective. The objective signal was measured when no sample was present, and this objective signal multiplied by 2.7 was subtracted from the sample signal. The value of 2.7 was calculated from $Objective\ signal \times \left(1 + R(\lambda_{ouut}) + R(\lambda_{in}) + R(\lambda_{out}) \times R(\lambda_{in})\right)$, where $R$ is the power reflection coefficient of the sample. This originates from the objective signal being reflected from the sample, plus the reflected laser causing the objective to fluoresce a second time when compared to the case of no sample present. We note this correction also removed the angular profile of the objective glow from our measurement. Lastly, we note that in real-space measurements the objective glow was orders of magnitude weaker than our sample signal, due to the sample signal originating from one small region in real space.

**Photoluminescence Quantum Yield (PLQY):** was estimated using a similar approach to that outlined by Frohna et al. [65]. Specifically, a calibrated light source (Ocean Optics HL-3P-INT-CAL) was coupled to an integrating sphere with a known spectral response (Thorlabs 2P3/M). The output port of this sphere was aligned with the objective lens of the Raman microscope and the spectrum recorded (allowing for relative radiometric calibration). The notch filter that normally removes the laser signal from the recording path was then removed, and the signal again recorded. The ratio of these signals gives the spectral response with and without the notch filter present. Secondly, the sphere was removed and the objective lens focused on a mirror (Thorlabs PF10-03-P10-10) with a known spectral response. The laser normally used to excite the sample was then shone on the mirror (at low power) and the signal recorded on the spectrometer. Finally, the mirror is removed and the incident power at the same position recorded. By accounting for the reflection strength of the mirror, the ratio of the number of photons incident on the mirror and the number recorded by the spectrometer can be calculated (i.e., an absolute calibration at one wavelength). Using the recorded lamp spectrum, it is then possible to work out an absolute calibration at any wavelength both with the notch filter present and removed. This calibration allows for PLQY measurement. We note that the relative radiometric response of the sphere was measured by sending a collimated white light beam into a second integrating sphere and recording the signal at the output port on an Ocean Insight Spectrometer FLAME-S-XR1. The sphere of interest was then coupled to the second integrating sphere and the collimated beam sent into the sphere of interest. The ratio of these two signals gives the spectral response of the integrating sphere used in measurements.

**Absorption measurements:** Absorption measurements were recorded on the NT&C system (a Nikon Eclipse Ti2 coupled to a Princeton Instruments Spectra Pro HRS-500 spectrometer and a PIXIS 256 camera). The light source was an Energetica LDLS™ laser-driven white light source output through a fibre with 100 μm core diameter. For transmission, incident light was collimated and then focused on the sample through a condenser lens with variable a-stop (almost fully closed), while for reflection incident light was focused in the centre of the objective back focal plane. In both cases this allowed light to be incident perpendicular to the sample. By appropriate choice of lenses for transmission, and by use of an iris for reflection, the illuminated area was reduced to approximately 4 $\mu m^2$ (i.e. smaller than the size of any gold flake). All recorded signals were passed through a 450 nm long pass filter (Thorlabs FELH0450) to prevent second-order effects. For transmission a bare quartz substrate was used as a reference (included in the modelling), while for reflection a mirror of known spectral response (Thorlabs PF10-03-P10-10) was employed as the reference (the Thorlabs reported spectral response was used). When calculating the signal per absorbed photon, the modelled absorption values were used (though we note extremely similar values are obtained using experimental values).

**Atomic Force Microscopy (AFM):** was used to measure sample thicknesses. A Bruker FastScan AFM was employed for all measurements in ScanAsyst mode. ScanAsyst-Fluid+ tips were used in all measurements. $0^{th}$ order flattening and $1^{st}$ order plane fits were applied to all results.

**Acknowledgements**

ARB and FK acknowledge support of SNSF Eccellenza Grant PCEGP2-194181. ARB acknowledges SNSF Swiss Postdoctoral Fellowship TMPFP2_217040 and thanks Valeria Vento and Christophe Galland for the use of a commercial monocrystalline 200 nm gold sample, and Franky Esteban Bedoya Lora for the use of the Ocean Optics spectrometer. ARE, FI and FJGA acknowledge funding from the European Research Council (Advanced Grant No. 789104-eNANO), the Spanish MICINN (PID2020–112625 GB-I00 and Severo Ochoa CEX2019-000910-S), the Catalan CERCA Program, and Fundaciós Cellex and Mir-Puig. JDC is a Sapere Aude research leader supported by VILLUM FONDEN (grant no. 16498) and Independent Research Fund Denmark (grant no. 0165-00051B). The Center for Polariton-driven Light—Matter Interactions (POLIMA) is funded by the Danish National Research Foundation (Project No. DNRF165). First-principles calculations were carried out at the Center for Computational Innovations at Rensselaer Polytechnic Institute.

# Supplemental Information: Quantum-mechanical effects in photoluminescence from thin crystalline gold films

A. R. Bowman[1], A. Rodríguez Echarri[2,3], F. Kiani[1], F. Iyikanat[2], T. V. Tsoulos[1], J. D. Cox[4,5], R. Sundararaman[6,7], F. Javier García de Abajo[2,8], and G. Tagliabue[1]*

1. Laboratory of Nanoscience for Energy Technologies (LNET), STI, École Polytechnique Fédérale de Lausanne (EPFL), Lausanne 1015, Switzerland
2. ICFO–Institut de Ciencies Fotoniques, The Barcelona Institute of Science and Technology, 08860 Castelldefels (Barcelona), Spain
3. MBI–Max-Born-Institut, 12489 Berlin, Germany
4. POLIMA–Center for Polariton-driven Light–Matter Interactions, University of Southern Denmark, Campusvej 55, DK-5230 Odense M, Denmark
5. Danish Institute for Advanced Study, University of Southern Denmark, Campusvej 55, DK-5230 Odense M, Denmark
6. Department of Materials Science & Engineering, Rensselaer Polytechnic Institute, 110 8th Street, Troy, New York 12180, USA
7. Department of Physics, Applied Physics, and Astronomy, Rensselaer Polytechnic Institute, 110 8th Street, Troy, New York 12180, USA
8. ICREA–Institució Catalana de Recerca i Estudis Avançats, Passeig Lluís Companys 23, 08010 Barcelona, Spain
*corresponding author: giulia.tagliabue@epfl.ch

**Supplemental Note 1 – Spatial resolution of flake emission**

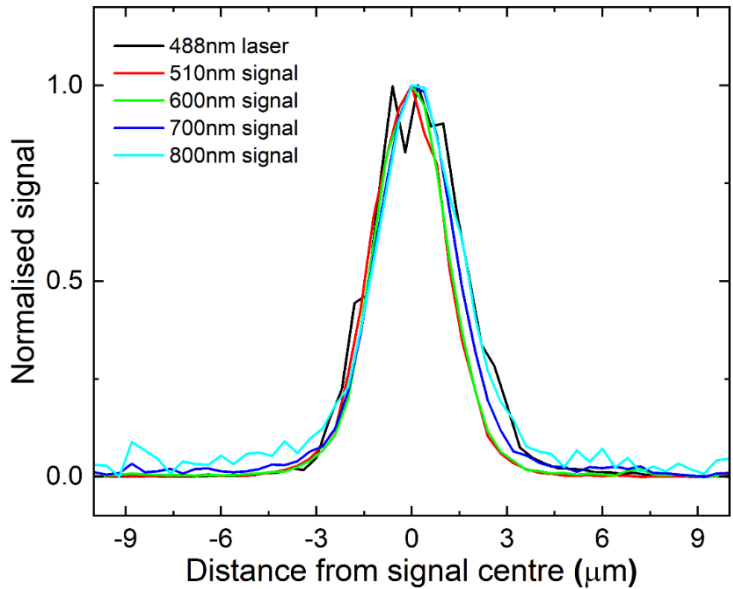

*Figure S1. In-plane spatial spread of the photoluminescence signal compared to the spread of the 488 nm laser beam for a flake of 113 nm thickness. The laser intensity used here is 0.079 mWµm$^{-2}$.*

In Figure S1, we superimpose the spatially resolved emission profile from a gold flake and the laser spot shape on the surface of the sample. The two are in strong agreement, showing that the signal only originates from the position where the laser is acting.

**Supplemental Note 2 – Comparing the emission from samples fabricated from different synthesis methods**

In Figure S2, we present the photoluminescence from a 110 nm gold flake synthesised by Kiani et al.'s method [1] and a 200 nm commercial flake on a mica substrate (note the gold is thick enough so the substrate does not affect the optical response) fabricated by Phasis [2], illuminated with the same laser intensity. The crystallographic orientation of exposed surface is (111) in both flakes . The signals are identical, confirming that our observations do not depend on the synthesis method.

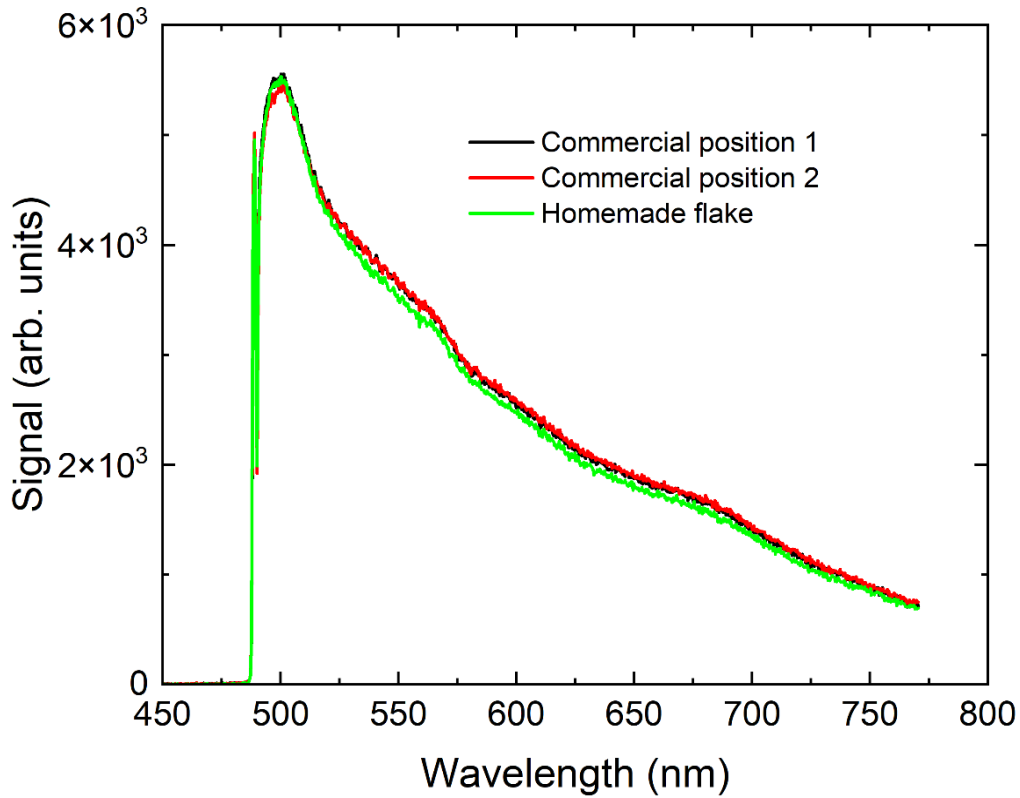

*Figure S2. Comparison of photoluminescence signals from a 110 nm flake fabricated by Kiani et al.'s method (green curve) and a 200 nm commercially available monocrystalline flake measured in two different positions (black and red curves) when excited by a 488 nm laser (0.042 mWμm$^{-2}$ intensity). We note that these signals are not radiometrically calibrated.*

**Supplemental Note 3 – Long-wavelength signals do not overlap in energy shift**

In Figure S3, we present the results from Figure 1a reformatted to show the signal as a function of energy shift relative to the incident laser intensity. When presented as energy shifts, the signals no longer overlap.

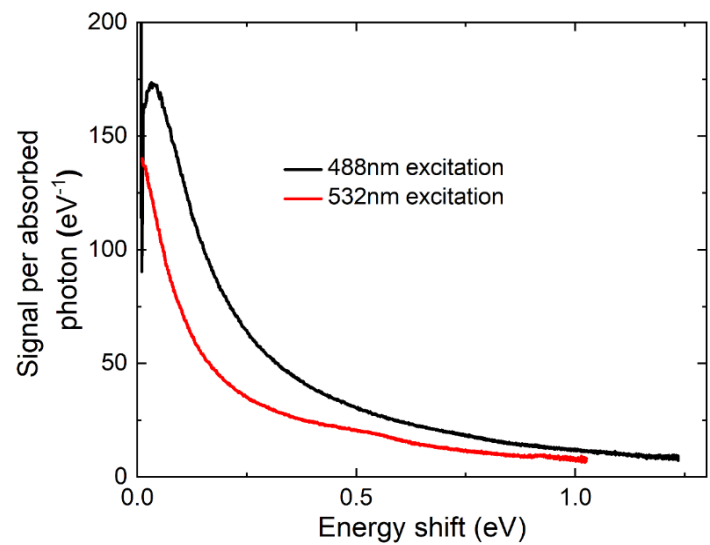

*Figure S3. Signal per absorbed photon for a 113 nm flake as a function of energy shift for 488 nm and 532 nm excitation wavelengths. The light intensity is 1.449 mWμm$^{-2}$/3.334 mWμm$^{-2}$ for 488 nm/532 nm excitation.*

**Supplemental Note 4 – Linear scaling of the local sample temperature with laser excitation intensity**

In Figure S4, we present the temperature extracted from the flake as a function of the absorbed laser power for 14 nm and 113 nm gold thicknesses. We find that thinner flakes reach higher temperatures for the same absorbed intensity, as expected from the increased thermal confinement. We note that there is larger uncertainty in the temperature measurement for the thinner flake as the cryostat sample stage also contributes a weak temperature-dependent Raman signal that competes more strongly with the signal emanating from thinner flakes. We measured this stage contribution separately and subtracted it from the gold signal, but some uncertainty remains.

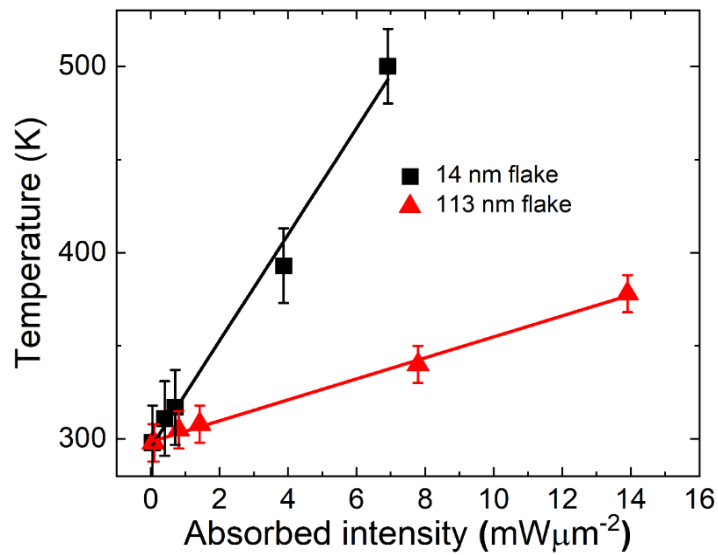

*Figure S4. Measured sample temperature as a function of absorbed laser power, as calculated from the measured laser power and the calculated absorption based on analysis in Supplemental Note 7. We show results for 14 nm and 113 nm flakes excited by 488 nm light.*

**Supplemental Note 5 – Scaling of photoluminescence quantum yield with excitation intensity**

We can define the external PLQY as

$$PLQY_{ext} = \frac{Photons\ out}{Photons\ absorbed} = \frac{Radiative\ recombination\ rate}{Total\ generation\ rate, G}.$$

As we are operating in steady state (at the excitation densities quoted more than 1 photon is being absorbed in the material every fs) and considering that the emitted photons we measure satisfy $\hbar\omega_{out} > \frac{\hbar\omega_{in}}{2}$, the generation rate is proportional to the total recombination rate, that is, $G \propto R$. Furthermore, in metals, at low excitation density (as is the case in continuous wave measurements), $R \propto n_{ex} = p_{ex}$, that is, the recombination rate scales linearly with the total number of excited electrons ($n_{ex}$), which is in turn equal to the number of excited holes ($p_{ex}$, to preserve charge neutrality, and assuming a minimal effect of charge traps). This is because, at low excitation density, both electron-electron and electron-phonon scattering are linear in the number of excited charge carriers and generation of additional electron-hole excited pairs thus scales linearly with the initial number of excited charge carriers.

In general, the radiative recombination rate is $\propto Ap_{ex}n_{ex} + Bn_{ex} + Cp_{ex}$, where the first term corresponds to recombination of excited electrons with excited holes, while the second and third terms correspond to the radiative recombination of excited electrons/holes with unexcited carriers, respectively. This expression involves constant coefficients $A$, $B$ and $C$. Combining the above expressions together, we find

$$PLQY_{ext} \propto AG + D,$$

where $G$ is proportional to the incident laser power and $D = B + C$. As we observe in Figure 1d, $PLQY_{ext}$ is constant as the laser power increases. Therefore, the recombination of excited charge carriers with unexcited charge carriers is dominant.

**Supplemental Note 6 – Electromagnetic theory of photoluminescence produced upon radiative recombination**

Electron-hole recombination events involve relatively small distances compared to the light wavelength, and therefore, we assimilate the emission to that of localized dipolar emitters. The actual strength and number of those emitters depend on the electron dynamics following laser irradiation, as discussed below. However, the electromagnetic aspects of the emission can be safely described by formulating a semi-analytical theory under the assumption of the localized-dipole-emitters model combined with the local response model (i.e., the involved materials are represented through their frequency-dependent local permittivities). Next, we present a derivation of the emission intensity produced by a dipolar emitter placed inside the metal film (see Fig. S5) based on the solution of Maxwell's equations, which is expressed in terms of the electromagnetic Green tensor. In this Supplemental Note we work in Gaussian units. We consider a thin metallic film with permittivity $\epsilon_\mathrm{m}$ and thickness $d$ that spans the region $z = 0$ to $z = d$ and interfaces a homogeneous superstrate (substrate) with permittivity $\epsilon_1$ ($\epsilon_3$), as schematically illustrated in Figure S5.

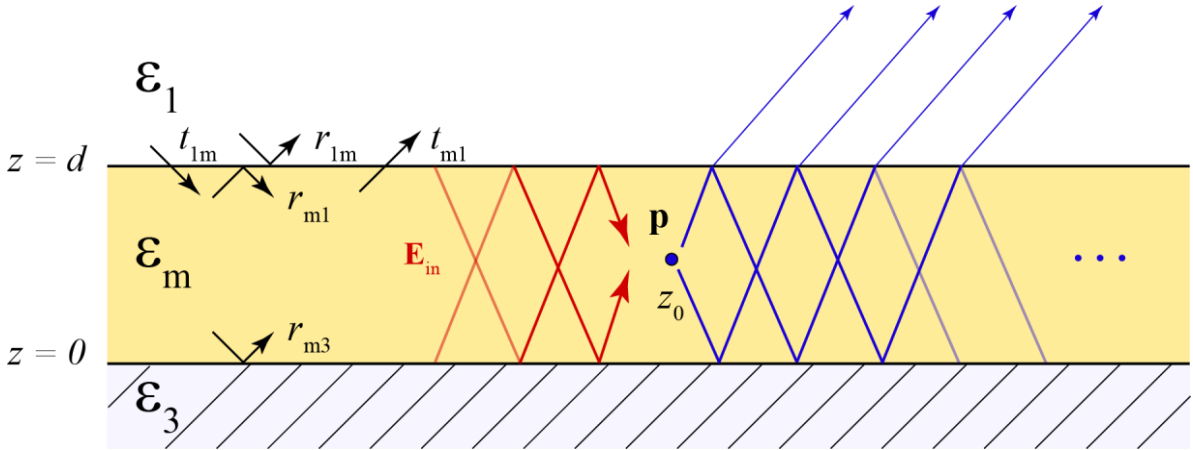

*Figure S5. Illustration of a dipole emitting inside a metallic film. A dipole $\mathbf{p}$, which is excited by an impinging field $\mathbf{E}^\mathrm{exc}$ (red lines), is placed at $z = z_0$ inside a film of thickness $d$ and permittivity $\epsilon_\mathrm{m}$ adjacent to homogeneous media with permittivity $\epsilon_1$ above and $\epsilon_3$ below. The dipole generates waves (blue lines) that undergo multiple reflections before they escape to the far field, where the emission is eventually detected.*

1. <u>Electromagnetic theory</u>

To compute the electric field outside the film, we follow a well-establish procedure [3] and start with the field produced at a position $\mathbf{r} = (\mathbf{R}, z)$, with $\mathbf{R} \equiv (x, y)$, by a dipole $\mathbf{p}$ located at $\mathbf{r}_0$ inside a homogeneous bulk metal [4],

$$\mathbf{E}^\mathrm{dip}(\mathbf{r}, \omega) = \frac{1}{\epsilon_\mathrm{m}}(k_\mathrm{m}^2 + \nabla \times \nabla) \, \mathbf{p} \frac{e^{ik_\mathrm{m}|\mathbf{r}-\mathbf{r}_0|}}{|\mathbf{r} - \mathbf{r}_0|},$$

where the wavenumber $k_\mathrm{m} = k\sqrt{\epsilon_\mathrm{m}}$ in the metal is expressed in terms of the vacuum wavenumber $k = \omega/c$ and $\omega$ is the emission frequency. Using the identity

$$\frac{e^{ik_\mathrm{m}|\mathbf{r}-\mathbf{r}_0|}}{|\mathbf{r}-\mathbf{r}_0|} = \int \frac{d^2\mathbf{Q}}{(2\pi)^2} \frac{2\pi i}{k_{\mathrm{m}z}} e^{i\mathbf{Q}\cdot(\mathbf{R}-\mathbf{R}_0)} e^{ik_{\mathrm{m}z}|z-z_0|},$$

in which we have introduced the in-plane wave vector $\mathbf{Q} = (Q_x, Q_y)$ and the normal component $k_{\mathrm{m}z} = \sqrt{k_\mathrm{m}^2 - Q^2}$ (with the square root taken such that $\mathrm{Im}\{k_{\mathrm{m}z}\} > 0$), along with the unit vectors $\hat{k}_\mathrm{m}^\pm = \frac{\mathbf{Q}\pm k_{\mathrm{m}z}\hat{z}}{k_\mathrm{m}}$, $\mathbf{e}_s = \frac{-Q_y\hat{x}+Q_x\hat{y}}{Q}$, and $\mathbf{e}_{\mathrm{p,m}}^\pm = \frac{\pm k_{\mathrm{m}z}\mathbf{Q}-Q^2\hat{z}}{Qk_\mathrm{m}}$ forming a complete and orthonormal set, the dipole field becomes

$$\mathbf{E}^{\mathrm{dip}}(\mathbf{r},\omega) = \frac{ik^2}{2\pi} \int \frac{d^2\mathbf{Q}}{k_{\mathrm{m}z}} \left[(\mathbf{e}_s \cdot \mathbf{p})\mathbf{e}_s + (\mathbf{e}_{\mathrm{p,m}}^\pm \cdot \mathbf{p})\mathbf{e}_{\mathrm{p,m}}^\pm\right] e^{i\mathbf{Q}\cdot(\mathbf{R}-\mathbf{R}_0)} e^{ik_{\mathrm{m}z}|z-z_0|}.$$

However, in the metal film, the generated electromagnetic s- and p-waves travelling up and down undergo reflections at the interfaces. Taking interface scattering into account, the electric field above the film (i.e., in medium 1 outside the metal film; see Fig. S5), where we measure the luminescence signal, is given by

$$\mathbf{E}_\mathrm{m}(\mathbf{r},\omega) = \frac{ik^2}{2\pi} \int \frac{d^2\mathbf{Q}}{k_{\mathrm{m}z}} e^{i\mathbf{Q}\cdot(\mathbf{R}-\mathbf{R}_0)} e^{ik_z|z-z_0|}$$
$$\times \left\{ \begin{array}{l} \mathbf{e}_\mathrm{p}^+ \left[ (\mathbf{e}_{\mathrm{p,m}}^+ \cdot \mathbf{p}) \frac{t_{\mathrm{p,m1}} e^{ik_{\mathrm{m}z}(d-z_0)}}{1 - r_{\mathrm{p,m1}} r_{\mathrm{p,m3}} e^{2ik_{\mathrm{m}z}d}} + (\mathbf{e}_{\mathrm{p,m}}^- \cdot \mathbf{p}) \frac{r_{\mathrm{p,m3}} t_{\mathrm{p,m1}} e^{ik_{\mathrm{m}z}(d+z_0)}}{1 - r_{\mathrm{p,m1}} r_{\mathrm{p,m3}} e^{2ik_{\mathrm{m}z}d}} \right] \\ \mathbf{e}_s (\mathbf{e}_s \cdot \mathbf{p}) \frac{t_{\mathrm{s,m1}}}{1 - r_{\mathrm{s,m1}} r_{\mathrm{s,m3}} e^{2ik_{\mathrm{m}z}d}} \left[ e^{ik_{\mathrm{m}z}(d-z_0)} + r_{\mathrm{s,m3}} e^{ik_{\mathrm{m}z}(d+z_0)} \right] \end{array} \right\}, \quad (S1)$$

where we introduce Fresnel transmission and reflection coefficients defined as [3]

$$r_{\mathrm{s,m1}} = \frac{k_{\mathrm{m}z} - k_{1z}}{k_{\mathrm{m}z} + k_{1z}}, \qquad t_{\mathrm{s,m1}} = \frac{2k_{\mathrm{m}z}}{k_{\mathrm{m}z} + k_{1z}},$$

$$r_{\mathrm{p,m1}} = \frac{\epsilon_1 k_{\mathrm{m}z} - \epsilon_\mathrm{m} k_{1z}}{\epsilon_1 k_{\mathrm{m}z} + \epsilon_\mathrm{m} k_{1z}}, \qquad t_{\mathrm{p,1m}} = \frac{2\sqrt{\epsilon_\mathrm{m}\epsilon_1} k_{\mathrm{m}z}}{\epsilon_1 k_{\mathrm{m}z} + \epsilon_\mathrm{m} k_{1z}}.$$

We can find an expression for the electric field in the far field (FF) in air that takes the form $\mathbf{E}^{\mathrm{FF}} = \mathbf{f}(\hat{\mathbf{r}})\frac{e^{ikr}}{r}$ by taking the limit $kr \to \infty$ in the expression of the electric field $\mathbf{E}_\mathrm{m}(\mathbf{r},\omega)$, where we apply the identity

$$\lim_{kr\to\infty} \int \frac{d^2\mathbf{Q}}{(2\pi)^2} \frac{2\pi i}{k_z} e^{i\mathbf{k}\cdot\mathbf{r}} g(\mathbf{Q}) = \left[ g(\mathbf{Q}) \frac{e^{ikr}}{r} \right]_{\mathbf{Q}=k\frac{\mathbf{R}}{r}}$$

for a given kernel $g(\mathbf{Q})$. The left-hand side of this equation has the same structure as Eq. (S1), from which we obtain an explicit expression of the kernel. We thus find

$$|f(\hat{r})|^2$$

$$= \left|e^{-ik_zd}\right|^2 \left|\frac{k^2 k_z}{k_{mz}}\right|^2 \left\{ \begin{array}{l} \left|(\mathbf{e}_{p,m}^+ \cdot \mathbf{p})\dfrac{t_{p,m1} e^{ik_{mz}(d-z_0)}}{1 - r_{p,m1}r_{p,m3}e^{2ik_{mz}d}} + (\mathbf{e}_{p,m}^- \cdot \mathbf{p})\dfrac{r_{p,m3} t_{p,m1} e^{ik_{mz}(d+z_0)}}{1 - r_{p,m1}r_{p,m3}e^{2ik_{mz}d}}\right|^2 \\ \left|(\mathbf{e}_s \cdot \mathbf{p})\dfrac{t_{s,m1}}{1 - r_{s,m1}r_{s,m3}e^{2ik_{mz}d}} \left[e^{ik_{mz}(d-z_0)} + r_{s,m3}e^{ik_{mz}(d+z_0)}\right]\right|^2 \end{array} \right\}.$$

We now assume incoherent dipoles of equal weight oriented along $x, y$, and $z$ directions with magnitudes $p_x = p_y = p_z \equiv p$. The corresponding projections of the unit vectors on Cartesian axes read

$$\mathbf{e}_s \cdot \hat{x} = -\frac{Q_y}{Q}, \qquad \mathbf{e}_s \cdot \hat{y} = \frac{Q_x}{Q}, \qquad \mathbf{e}_s \cdot \hat{z} = 0,$$

and

$$\mathbf{e}_{p,m}^\pm \cdot \hat{x} = \pm\frac{Q_x k_{mz}}{Q k_m}, \qquad \mathbf{e}_{p,m}^\pm \cdot \hat{y} = \pm\frac{Q_y k_{mz}}{Q k_m}, \qquad \mathbf{e}_{p,m}^\pm \cdot \hat{z} = -\frac{Q}{k_m}.$$

This allows us to write the total far-field intensity and averaging over the three Cartesian dipole orientations as

$$|f(\hat{r}, z_0)|^2 = \left|e^{-ik_zd}\right|^2 k^4 p^2 \left\{ \left|\frac{k_{mz}}{k_m}\right|^2 D_p^- + D_s^+ + \left|\frac{Q}{k_m}\right|^2 D_p^+ \right\},$$

where

$$D_\sigma^\pm(\theta, z_0) = \left|\frac{t_{\sigma,1m}}{1 - r_{\sigma,m1} r_{\sigma,m3} e^{2ik_{mz}d}} \left[e^{ik_{mz}(d-z_0)} \pm r_{\sigma,m3} e^{ik_{mz}(d+z_0)}\right]\right|^2$$

depends explicitly on the position $z_0$ and the emission angle $\theta$, and we have used the identity $\left|\frac{k_z t_{\sigma,m1}}{k_{mz}}\right|^2 = |t_{\sigma,1m}|^2$.

We are interested in the flux of photons emitted into the far field, which we obtain from the Poynting vector. The latter can be calculated as $\mathbf{S} = \left(\frac{c}{4\pi}\right)\left(\mathbf{E}^{FF} + \text{c.c.}\right) \times \left(\mathbf{H}^{FF} + \text{c.c.}\right)$, in which the magnetic field is obtained from Faraday's law, $\mathbf{H}^{FF} = -\left(\frac{i}{k}\right)\nabla \times \mathbf{E}^{FF}$. The time-averaged power flux is given by

$$\langle \hat{r} \cdot \mathbf{S} \rangle_{time} = \frac{c}{2\pi r^2} |f(\hat{r}, z_0)|^2,$$

where $\hat{r} = \frac{\mathbf{k}}{k}$ is the unit vector indicating the emission direction. From here, we can write the flux of photons emitted per unit of solid angle as

$$\frac{d\phi(\theta, z_0)}{d\Omega} = \frac{r^2}{\hbar\omega}\langle \hat{r}\cdot \mathbf{S}\rangle_{time} = \frac{c}{2\pi\hbar\omega}|\mathbf{f}(\hat{r}, z_0)|^2$$

$$= \frac{1}{2\pi\hbar k} k^4 p^2 \left\{ \left|\frac{k_{mz}}{k_m}\right|^2 D_p^-(\theta, z_0) + D_s^+(\theta, z_0) + \left|\frac{Q}{k_m}\right|^2 D_p^+(\theta, z_0) \right\}.$$

Integrating over emission solid angle (including azimuthal angle), we obtain the photoluminescence intensity emanating from a given induced dipole position at depth $z_0$ as

$$\phi(z_0) = \frac{k^3 p^2}{\hbar} \int_0^{\theta_{max}} d\theta \left\{ \left|\frac{k_{mz}}{k_m}\right|^2 D_p^-(\theta, z_0) + D_s^+(\theta, z_0) + \left|\frac{Q}{k_m}\right|^2 D_p^+(\theta, z_0) \right\} \sin(\theta), \qquad (S2)$$

where the integral extends over the angles allowed by the numerical aperture determined by the acceptance angle $\theta_{max}$. Finally, we define the dipole strength per unit volume as $p^2 = \tilde{p}(z_0)^2 dV$ (with a specific dependence on $z_0$), allowing us to write the photon flux per unit area as

$$\tilde{\phi}(z_0) = \frac{k^3 \tilde{p}(z_0)^2 dz_0}{\hbar} \int_0^{\theta_{max}} d\theta \left\{ \left|\frac{k_{mz}}{k_m}\right|^2 D_p^-(\theta, z_0) + D_s^+(\theta, z_0) + \left|\frac{Q}{k_m}\right|^2 D_p^+(\theta, z_0) \right\} \sin(\theta).$$

2. The role of electron transport

The total photon emission is then obtained by accumulating the incoherent sum of the emission produced by dipoles distributed along positions $z_0$ across the depth of the film. At each position inside the metal film, the emission is proportional to the dipole strength per unit volume $\tilde{p}(z_0)^2$, which is in turn dependent on incidence and emission photon energies. The dipole strength depends on the number of excited charges (normalized per incident photon) and their transport to the positions at which they produce luminescence, away from the locations where the excitation takes place. The actual mechanisms of transport are complex and follow an intricate dynamics in which a cascade of electron-hole pairs is generated, assisted by electron-electron interactions, as well as scattering by phonons and impurities in general and diffusion. Transport and dynamics are thus intimately related, as captured by the general expression

$$\tilde{p}(z_0) = \int_0^d dz_0'\, R(z_0, z_0', \omega_{in}, \omega_{out})\, |\mathbf{E}^{exc}(z_0', \omega_{in})|^2,$$

which reflects the fact that the primary excitation at a location $z_0'$ is proportional to the near-field intensity $|\mathbf{E}^{exc}(z_0', \omega_{in})|^2$ associated with the incident light (including its scattering by the planar interfaces; see Figure S5 and also an explicit expression below), as well as a nonlocal dynamics function $R(z_0, z_0', \omega_{in}, \omega_{out})$ (we denote the outgoing photon frequency as $\omega = \omega_{out}$ for brevity in the above

derivation). In what follows, we show that transport plays a negligible role by considering two limiting cases:

(1) *Local PL model*. In this limit, we consider that the excitation and emission processes take place at the same position, such that transport can be neglected. The emission dipole strength is then given without loss of generality by

$$\tilde{p}(z_0) = G(\omega_{\text{in}}, \omega_{\text{out}}) |\mathbf{E}^{\text{exc}}(z_0, \omega_{\text{in}})|^2, \quad \text{(S3a)}$$

which is just proportional to the near-field intensity $|\mathbf{E}^{\text{exc}}(z_0, \omega_{\text{in}})|^2$ at the emission position $z_0$. Here, the dynamics function $R(z_0, z_0', \omega_{\text{in}}, \omega_{\text{out}})$ is simplified under the local approximation (i.e., only parameters $z_0 = z_0'$ contribute), and further neglecting interface effects in the energy conversion process. We conclude that $p^2(z_0)$ is linearly dependent on a position-independent conversion function $G(\omega_{\text{in}}, \omega_{\text{out}})$ that describes the local electron dynamics and the radiative recombination processes. In particular, $G(\omega_{\text{in}}, \omega_{\text{out}})$ can be computed just by considering the properties of the bulk metal, as obtained from the *ab-initio* calculations that we discuss below.

(2) *Maximally delocalized PL model*. In the opposite limit, any excitation inside the metal contributes to generate an emitting dipole at all film positions, such that

$$\tilde{p}^2 = G(\omega_{\text{in}}, \omega_{\text{out}}) \frac{1}{d} \int_0^d dz_0' \, |\mathbf{E}^{\text{exc}}(z_0', \omega_{\text{in}})|^2 \quad \text{(S3b)}$$

becomes position-independent. Here, $G(\omega_{in}, \omega_{out})$ is the same function as in model (1).

We show in the main text that model (1) (local PL) agrees excellently with our measurements, whereas model (2) produces strong discrepancies.

3. <u>Position integrated photoluminescence</u>

In both transport models described by Eqs. (S3a) and (S3b), we calculate the photon flux per unit area by integrating the signal across the film as

$$\widetilde{PL}_{\text{external}} = \int_0^d \tilde{\phi}(z_0).$$

Combining this expression with Eqs. (S2) and (S3), we find

$$\widetilde{PL}_{\text{external}} = \frac{k^3}{\hbar} G(\omega_{\text{in}}, \omega_{\text{out}}) \int_0^d dz_0 \, |\mathbf{E}^{\text{exc}}(z_0, \omega_{\text{in}})|^2 \int_0^{\theta_{max}} d\theta \left\{ \left|\frac{k_{\text{mz}}}{k_{\text{m}}}\right|^2 D_{\text{p}}^-(\theta, z_0) + D_{\text{s}}^+(\theta, z_0) \right.$$

$$\left. + \left|\frac{Q}{k_{\text{m}}}\right|^2 D_{\text{p}}^+(\theta, z_0) \right\} \sin(\theta) \quad \text{(S4a)}$$

in the local PL model, and

$$\widetilde{PL}_{\text{external}} = \frac{k^3}{\hbar} G(\omega_{\text{in}}, \omega_{\text{out}}) \left( \int_0^d dz_0' \; |\mathbf{E}^{\text{exc}}(z_0', \omega_{\text{in}})|^2 \right) \int_0^d dz_0 \int_0^{\theta_{max}} d\theta \left\{ \left|\frac{k_{\text{mz}}}{k_{\text{m}}}\right|^2 D_{\text{p}}^-(\theta, z_0) \right.$$

$$\left. + D_{\text{s}}^+(\theta, z_0) + \left|\frac{Q}{k_{\text{m}}}\right|^2 D_{\text{p}}^+(\theta, z_0) \right\} \sin(\theta) \qquad \text{(S4b)}$$

in the maximally delocalized PL model. These expression can be related to the absorption power density (power per unit volume) inside the metal film by using the expression [5]

$$P(z_0) = \frac{\omega_{\text{in}}}{2\pi} \text{Im}\{\epsilon_m(\omega_{\text{in}})\} |\mathbf{E}^{\text{exc}}(z_0, \omega_{\text{in}})|^2.$$

In particular, Eqs. (S3a) and (S3b) become

$$\widetilde{PL}_{\text{external}} = \frac{2\pi k^3}{\omega_{\text{in}} \hbar \text{Im}\{\epsilon_m(\omega_{\text{in}})\}} G(\omega_{\text{in}}, \omega_{\text{out}}) \int_0^d dz_0 \; P(z_0) \int_0^{\theta_{max}} d\theta \left\{ \left|\frac{k_{\text{mz}}}{k_{\text{m}}}\right|^2 D_{\text{p}}^-(\theta, z_0) \right.$$

$$\left. + D_{\text{s}}^+(\theta, z_0) + \left|\frac{Q}{k_{\text{m}}}\right|^2 D_{\text{p}}^+(\theta, z_0) \right\} \sin(\theta) \qquad \text{(S5a)}$$

and

$$\widetilde{PL}_{\text{external}} = \frac{2\pi k^3}{\omega_{\text{in}} \hbar \text{Im}\{\epsilon_m(\omega_{\text{in}})\}} G(\omega_{\text{in}}, \omega_{\text{out}}) P \int_0^d dz_0 \int_0^{\theta_{max}} d\theta \left\{ \left|\frac{k_{\text{mz}}}{k_{\text{m}}}\right|^2 D_{\text{p}}^-(\theta, z_0) + D_{\text{s}}^+(\theta, z_0) \right.$$

$$\left. + \left|\frac{Q}{k_{\text{m}}}\right|^2 D_{\text{p}}^+(\theta, z_0) \right\} \sin(\theta), \qquad \text{(S5b)}$$

respectively, where $P = \int_0^d dz_0 \; P(z_0)$ is the total absorbed power per unit of surface area.

4. <u>Near-field excitation field intensity</u>

For a plane wave normally impinging the metal surface (from medium $\epsilon_1$) at wavelength $\lambda_{\text{in}} = 2\pi c/\omega_{\text{in}}$ with incident electric-field amplitude $E_{in}$, the electric field inside the slab is

$$\mathbf{E}_\sigma^{\text{exc}}(z_0, \omega_{\text{in}}) = E_{\text{in}} \frac{t_{\sigma 1m}^{(\text{in})}}{1 - r_{\sigma m1}^{(\text{in})} r_{\sigma m3}^{(\text{in})} e^{2ik_{\text{mz}}^{(\text{in})} d}} \left[ \mathbf{e}_\sigma^- \, e^{-ik_{\text{mz}}^{(\text{in})}(d-z_0)} + \mathbf{e}_\sigma^+ r_{\sigma m3}^{(\text{in})} e^{ik_{\text{mz}}^{(\text{in})}(d+z_0)} \right] = E_{\text{in}} C(z_0),$$

where $k_{\text{mz}}^{(\text{in})} = 2\pi/\lambda_{\text{in}} \sqrt{\epsilon_{\text{m}}(\lambda_{\text{in}})}$, while $\sigma = \{\text{s}, \text{p}\}$ depends on the light polarization. In particular, under the conditions of the present experiment, we can approximate the light incidence to be normal to the film, and then, we have $r_{\text{p}}^{(\text{in})} = -r_{\text{s}}^{(\text{in})}$ and $t_{\text{p}}^{(\text{in})} = t_{\text{s}}^{(\text{in})}$, so that the intensity is independent of the incident light polarization.

### 5. Connection with DFT calculations: equating dipole strength with internal photoluminescence

For ease of notation, we make a connection between the emission dipole strength $p^2$ and direct optical transitions, noting that the same analysis can be carried out for phonon-assisted and pre-scattered transitions and we obtain analogous prefactors. Within the DFT formalism, we obtain the total density of dipole strength (normalized per unit volume and emission frequency $\omega = \omega_{out}$) due to direct electron-hole-recombination transitions as

$$\frac{p^2_{\text{direct}}}{dVd\omega_{\text{out}}} = \frac{\hbar e^2}{m_e^2 \omega_{\text{out}}^2} \int_{BZ} \frac{g_s d\vec{k}}{(2\pi)^3} \sum_{n',n} f_{\vec{k},n'}(1 - f_{\vec{k},n}) \delta(\varepsilon_{\vec{k}n'} - \varepsilon_{\vec{k}n} - \hbar\omega_{\text{out}}) \left|\vec{p}^{\,\vec{k}}_{n,n'}\right|^2.$$

Here, $e$ is the charge of an electron, $m_e$ is the mass of an electron, $g_s = 2$ is the spin degeneracy, $f_{\vec{k},n}$ is the probability that a one-electron state with wave vector $\vec{k}$ and band index $n$ is occupied (i.e., the occupation factor), $\varepsilon_{\vec{k}n}$ the energy eigenvalue of the relevant state, $\hbar$ Planck's constant divided by $2\pi$ and $\vec{p}^{\,\vec{k}}_{n,n'}$ matrix elements of the momentum operator. See Supplemental Note 8 and Brown *et al*. [6] for more details. Furthermore, we can write the spontaneous emission rate of photons per unit $\omega_{\text{out}}$ and volume from an isotropic/cubic crystal (see Supplemental Note 8 for further details) as [7]

$$\Gamma_{\text{direct}}(\omega_{\text{in}}, \omega_{\text{out}}) = \frac{4e^2 \omega_{\text{out}} \text{Re}\{\sqrt{\epsilon_m(\omega_{\text{out}})}\}}{m_e^2 c^3} \int_{BZ} \frac{g_s d\vec{k}}{(2\pi)^3} \sum_{n',n} f_{\vec{k},n'}(1 - f_{\vec{k},n}) \delta(\varepsilon_{\vec{k}n'} - \varepsilon_{\vec{k}n} - \hbar\omega) \left|\vec{p}^{\,\vec{k}}_{n,n'}\right|^2$$

and therefore

$$\frac{\tilde{p}^2_{direct}}{d\omega_{out}} = \frac{\hbar}{4k^3 \text{Re}\{\sqrt{\epsilon_m(\omega_{\text{out}})}\}} \Gamma_{\text{direct}}(\omega_{\text{in}}, \omega_{\text{out}}, z_0). \tag{S6}$$

The same prefactors are obtained for phonon-assisted or pre-scattered luminescence. Here we are specifically interested in emission beyond that of a black body at equilibrium. As discussed in Supplemental Note 8, in this case we can derive a spontaneous emission rate per absorbed photon, $PL_{\text{internal}}(\omega_{\text{in}}, \omega_{\text{out}})$ for the luminescence due to laser excitation, and the number of locally absorbed photons per unit volume is $\frac{P(z_0)}{\hbar \omega_{\text{in}}}$. Thin film effects are accounted for in the above dipole formalism, so we consider the bulk $PL_{\text{internal}}(\omega_{\text{in}}, \omega_{\text{out}})$ in a similar way as in the context of semiconductor PL, although we caution that the physical meaning of internal photon emission is compromised by losses and the deep-subwaverlength thickness of the films in the present work. In the case of minimal charge redistribution prior to luminescence we can state that the local laser-stimulated spontaneous emission rate is

$$\Gamma = PL_{\text{internal}}(\omega_{\text{in}}, \omega_{\text{out}}) \frac{P(z_0)}{\hbar \omega_{\text{in}}}$$

while in the maximally delocalized PL model the laser-stimulated spontaneous emission rate

$$\Gamma = PL_{\text{internal}}(\omega_{\text{in}}, \omega_{\text{out}}) \frac{1}{d} \int \frac{P(z_0{'})}{\hbar \omega_{\text{in}}} dz_0'.$$

Taking equations S3a and S6 allows us to equate

$$\tilde{p}_{direct}^2(z_0) = G(\omega_{\text{in}}, \omega_{\text{out}}) \, |\mathbf{E}^{\text{exc}}(z_0, \omega_{\text{in}})|^2$$
$$= \frac{\hbar}{4k^3 \text{Re}\{\sqrt{\epsilon_m(\omega_{\text{out}})}\}} PL_{\text{internal}}(\omega_{\text{in}}, \omega_{\text{out}}) \frac{P(z_0)}{\hbar \omega_{\text{in}}} d\omega_{\text{out}}$$

and from equation S3b

$$\tilde{p}_{direct}^2 = G(\omega_{\text{in}}, \omega_{\text{out}}) \frac{1}{d} \int_0^d dz_0' \, |\mathbf{E}^{\text{exc}}(z_0', \omega_{\text{in}})|^2$$
$$= \frac{\hbar}{4k^3 \text{Re}\{\sqrt{\epsilon_m(\omega_{\text{out}})}\}} PL_{\text{internal}}(\omega_{\text{in}}, \omega_{\text{out}}) \frac{1}{d} \int \frac{P(z_0{'})}{\hbar \omega_{\text{in}}} dz_0' \, d\omega_{out}.$$

In both cases this gives

$$G(\omega_{\text{in}}, \omega_{\text{out}}) = \frac{\hbar PL_{\text{internal}}(\omega_{\text{in}}, \omega_{\text{out}}) \frac{\omega_{\text{in}}}{2\pi} \text{Im}\{\epsilon_m(\omega_{\text{in}})\} d\omega_{\text{out}}}{4k^3 \text{Re}\{\sqrt{\epsilon_m(\omega_{\text{out}})}\} \hbar \omega_{\text{in}}}.$$

Substituting into Eqs. (S5a) and (S5b), and dividing by $d\omega_{\text{out}}$ to give the recorded signal per unit area, per unit $\omega_{\text{out}}$, we obtain

$$PL_{\text{external}} = \frac{PL_{\text{internal}}}{4\text{Re}\{\sqrt{\epsilon_m(\omega_{\text{out}})}\}} \int_0^d dz_0 \left(\frac{P(z_0)}{\hbar \omega_{\text{in}}}\right) \int_0^{\theta_{max}} d\theta \left\{ \left|\frac{k_{\text{m}z}}{k_{\text{m}}}\right|^2 D_p^-(\theta, z_0) + D_s^+(\theta, z_0) \right.$$
$$\left. + \left|\frac{Q}{k_{\text{m}}}\right|^2 D_p^+(\theta, z_0) \right\} \sin(\theta)$$

for the local PL model and

$$PL_{\text{external}} = \frac{PL_{\text{internal}}}{4\text{Re}\{\sqrt{\epsilon_m(\omega_{\text{out}})}\}} \left(\frac{P}{\hbar\omega_{\text{in}}}\right) \int_0^d dz_0 \int_0^{\theta_{max}} d\theta \left\{ \left|\frac{k_{mz}}{k_m}\right|^2 D_p^-(\theta, z_0) + D_s^+(\theta, z_0) \right.$$
$$\left. + \left|\frac{Q}{k_m}\right|^2 D_p^+(\theta, z_0) \right\} \sin(\theta)$$

for the maximally delocalised model. We note the factor of 4 in the denominator originates from the definition of $PL_{\text{internal}}$ (in accordance with the semiconductor luminescence community) rather than the electromagnetic derivation.

6. <u>Simplifying notation</u>

We can re-write

$$P(z_0) = \frac{\omega_{\text{in}}}{2\pi} \text{Im}\{\epsilon_m(\omega_{\text{in}})\} |\mathbf{E}^{\text{exc}}(z_0, \omega_{\text{in}})|^2 = \frac{\omega_{\text{in}}}{2\pi} \text{Im}\{\epsilon_m(\omega_{\text{in}})\}|E_{\text{in}}|^2|C(z_0)|^2 = f_{\text{abs}}(z_0, \lambda_{\text{in}})I_{\text{in}},$$

where $I_{\text{in}}$ is the incident intensity that can be experimentally recorded (via power meter and spot-size measurements) and $f_{\text{abs}}$ describes the number of photons absorbed at $z_0$ (per unit length). We have changed from using $\omega$ to wavelength $\lambda$ as experiments are recorded as a function of wavelength. We group all other terms, which describe energy loss prior to photons escaping the film, as

$$f_{\text{emit}}(z_0, \lambda_{\text{out}}) = \int_0^{\theta_{max}} d\theta \left\{ \left|\frac{k_{mz}}{k_m}\right|^2 D_p^- + D_s^+ + \left|\frac{Q}{k_m}\right|^2 D_p^+ \right\} \left(\frac{\sin(\theta)}{4Re\{\sqrt{\epsilon_m(\omega_{\text{out}})}\}}\right).$$

This allows us to write the two models as

$$PL_{\text{external}} = PL_{\text{internal}} \left(\frac{I_{\text{in}}}{\hbar\omega_{\text{in}}}\right) \int_0^d dz_0 \, f_{\text{abs}}(z_0, \lambda_{\text{in}}) \, f_{\text{emit}}(z_0, \lambda_{\text{out}})$$

for local PL and

$$PL_{\text{external}} = PL_{\text{internal}}(\omega_{\text{in}}, \omega_{\text{out}}) \left(\frac{I_{in}}{\hbar\omega_{\text{in}}}\right) \int_0^d dz_0' \, f_{\text{abs}}(z_0', \lambda_{\text{in}}) \int_0^d dz_0 \, f_{\text{emit}}(z_0, \lambda_{\text{out}})$$

for maximally delocalised PL. Finally, as an example, in Figure S6 we plot $\int dz_0 f_{abs}(z_0, \lambda_{\text{in}}) f_{\text{emit}}(z_0, \lambda_{\text{out}})$ for an 88 nm flake with 488 nm excitation and NA=0.7.

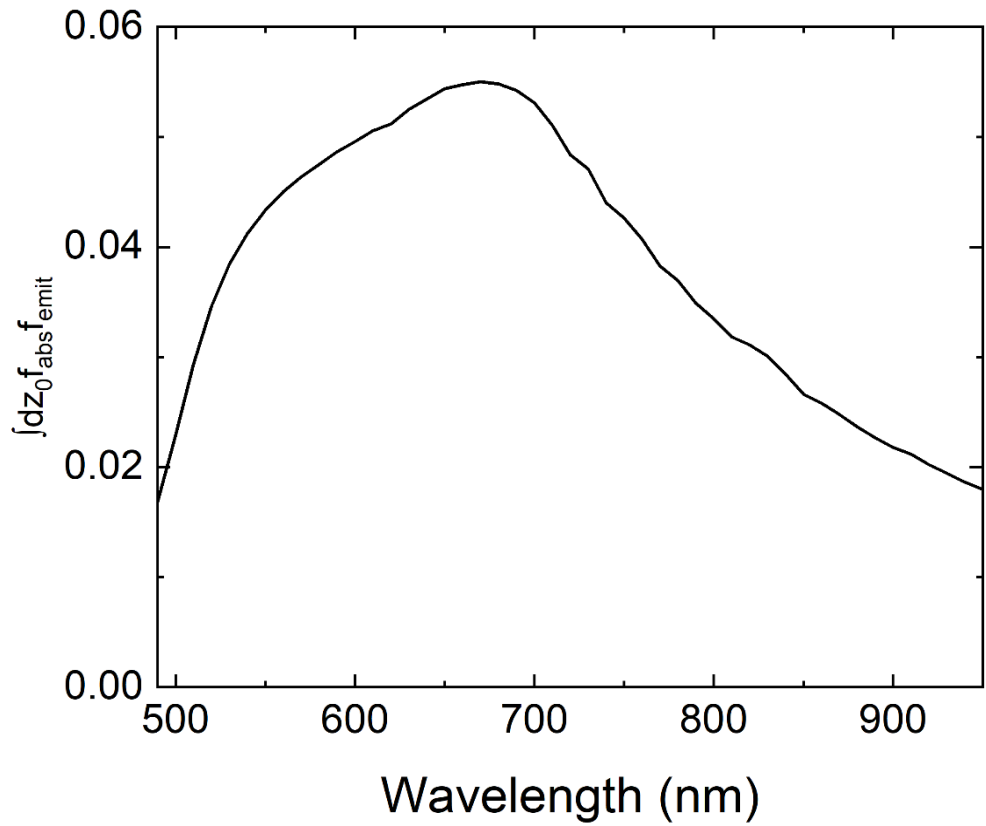

*Figure S6.* $\int dz_0 f_{abs}(z_0, \lambda_{in}) f_{emit}(z_0, \lambda_{out})$ *as a function of wavelength for an 88 nm flake when exciting at 488 nm and observing with a microscope objective NA=0.7.*

**Supplemental Note 7 – Measuring flake absorption with sample thickness**

1. Light absorption model

We consider gold on top of glass, so

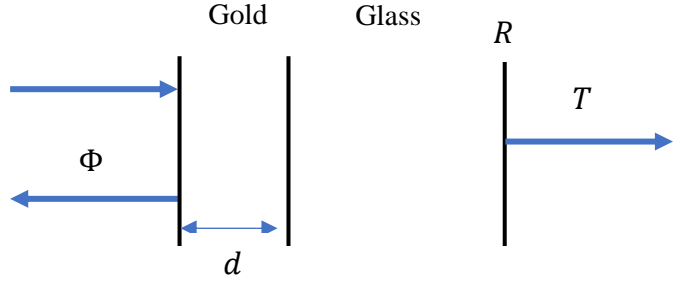

We consider that there is interference in gold, but all coherence is lost while light travels through the glass layer. Furthermore, we only write the formula for s polarization as, for light incident perpendicular on the sample, the two polarisations give the same result. If we consider gold has power transmission and reflection coefficients $R_L, R_R, T_{L \to R}$ and $T_{R \to L}$ (where L and R correspond to the left and right sides of gold), we can derive

$$T = \frac{(1-R)T_{L \to R}}{1-R_R R} \text{ and } \Phi = R_L + \frac{T_{R \to L}T_{L \to R}R}{1-R_R R}.$$

We can state

$$R_L = \left|\frac{-r_{s,m1}e^{-ik_m d} + r_{s,m3}e^{ik_m d}}{e^{-ik_m d} - r_{s,m1}r_{s,m3}e^{ik_m d}}\right|^2, R_R = \left|\frac{-r_{s,m3}e^{-ik_m d} + r_{s,m1}e^{ik_m d}}{e^{-ik_m d} - r_{s,m1}r_{s,m3}e^{ik_m d}}\right|^2, T_{L \to R}$$
$$= \frac{n_3}{n_1}\left|\frac{t_{s,m1out}t_{s,m3out}}{e^{-ik_m d} - r_{s,m1}r_{s,m3}e^{ik_m d}}\right|^2, \text{ and } T_{R \to L} = \frac{n_1}{n_3}\left|\frac{t_{s,m1}t_{s,m3}}{e^{-ik_m d} - r_{s,m1}r_{s,m3}e^{ik_m d}}\right|^2$$

where $n_3$ is the refractive index of quartz. We can write (for normal incidence)

$$t_{s,m1out} = \frac{2n_1}{n_1 + n_m}, t_{s,m3out} = \frac{2n_m}{n_m + n_3}, t_{s,m3} = \frac{2n_3}{n_m + n_3}$$

and

$$R = \left|\frac{n_1 - n_3}{n_1 + n_3}\right|^2.$$

Assuming that there is no absorption in the quartz, we can say the total absorption in the gold is

$$Abs = 1 - T - \Phi.$$

We note that this gives extremely similar values of absorption to the cases modelled with no substrate.

2. Experimental results

We present the reflection and transmission of a thin (14 nm) and moderately thick (47 nm) gold flake in Figures S7a and b respectively. In both cases we fit this data with the model presented in the previous section using McPeak's optical constants [8] (noting that other optical constants give similar results). The only free parameter in this fitting is the thickness of the flake. We present a comparison of this fitted thicknesses with that measured via atomic force microscopy (AFM) in Figure S7c, which shows strong agreement. This demonstrates that for both thin and thick flakes, optical constants in the literature are sufficient to understand the absorption properties of these flakes. We note that no additional factors were needed to model thin flakes, in agreement with Großmann et al. [9].

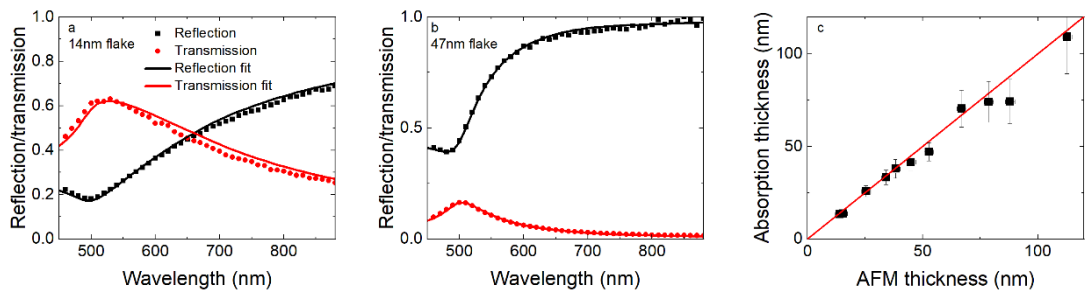

*Figure S7. a) and b) present data and fits of sample reflection and transmission for 14 nm and 47 nm flake respectively. c) Thicknesses extracted from absorption data relative to those measured from atomic force microscopy (AFM), with the red line showing the expected agreement. The legend in a) also applies to b).*

**Supplemental Note 8 – Modelling PL from DFT**

1. <u>Post-scattered carriers</u>

In first-principles calculations, the imaginary part of the dielectric function for direct and phonon-assisted transitions is calculated using (we denote the outgoint photon frequeny as $\omega = \omega_{out}$ for brevity in the following derivation):

$$\text{Im}\left(\epsilon_{m,direct}(\omega)\right) = \frac{4\pi^2 e^2}{m_e^2 \omega^2} \int_{BZ} \frac{g_s d\vec{k}}{(2\pi)^3} \sum_{n',n} (f_{\vec{k},n} - f_{\vec{k},n'}) \delta(\varepsilon_{\vec{k}n'} - \varepsilon_{\vec{k}n} - \hbar\omega) \left|\vec{p}_{n,n'}^{\vec{k}}\right|^2.$$

$$\text{Im}\left(\epsilon_{(m,phonon)}(\omega)\right)$$

$$= \frac{4\pi^2 e^2}{m_e^2 \omega^2} \int_{BZ} \frac{g_s d\vec{k}' d\vec{k}}{(2\pi)^6} \sum_{n'n\alpha\pm} (f_{\vec{k}n} - f_{\vec{k}'n'}) \left(n_{\vec{k}'-\vec{k},\alpha} + \frac{1}{2} \mp \frac{1}{2}\right) \delta(\varepsilon_{\vec{k}'n'} - \varepsilon_{\vec{k}n} - \hbar\omega$$

$$\mp \hbar\omega_{\vec{k}'-\vec{k},\alpha}) \left|\sum_{n_1} \left(\frac{g_{\vec{k}'n',\vec{k}n_1}^{\vec{k}'-\vec{k},\alpha} \vec{p}_{n_1 n}^{\vec{k}}}{\varepsilon_{\vec{k}n_1} - \varepsilon_{\vec{k}n} - \hbar\omega + i\eta} + \frac{\vec{p}_{n'n_1}^{\vec{k}'} g_{\vec{k}'n_1,\vec{k}n}^{\vec{k}'-\vec{k},\alpha}}{\varepsilon_{\vec{k}'n_1} - \varepsilon_{\vec{k}n} \mp \hbar\omega_{\vec{k}'-\vec{k},\alpha} + i\eta}\right)\right|^2$$

Here $e$ is the charge of an electron, $m_e$ the mass of an electron, $g_s$ is the spin degeneracy factor, $f_{\vec{k},n}$ the probability a state with wavevector $\vec{k}$ and band index $n$ is occupied (the occupation factor), $\varepsilon_{\vec{k}n}$ the energy eigenvalue of the relevant state, $\hbar$ Planck's constant divided by $2\pi$ and $\vec{p}_{n,n'}^{\vec{k}}$ matrix elements of the momentum operator, $n_{\vec{k}'-\vec{k},\alpha}$ the phonon occupation of the relevant state, $g_{\vec{k}'n',\vec{k}n_1}^{\vec{k}'-\vec{k},\alpha}$ the electron-phonon matrix elements and $\eta$ a small value to aid with calculation. See Brown et al. for more discussion of defined terms [6]. This can be further histogrammed by $\epsilon$ to compute hot carrier distributions, as implemented in JDFTx [10].

We present a derivation of PL for the case of direct transitions, and present the final equation for direct and phonon assisted transitions. We can write the spontaneous emission rate of photons per unit $\omega$ and volume from an isotropic/cubic crystal as [7]

$$\Gamma(\omega) = \frac{4e^2 \omega}{m_e^2 c^3} \text{Re}\{\sqrt{\epsilon_m(\omega)}\} \int_{BZ} \frac{g_s d\vec{k}}{(2\pi)^3} \sum_{n',n} f_{\vec{k},n'}(1 - f_{\vec{k},n}) \delta(\varepsilon_{\vec{k}n'} - \varepsilon_{\vec{k}n} - \hbar\omega) \left|\vec{p}_{n,n'}^{\vec{k}}\right|^2 \quad (S7)$$

where $\text{Re}\{\sqrt{\epsilon_m(\omega)}\}$ is the local photonic density of states. To calculate luminescence from a perturbed electron distribution, let $f_{\vec{k},n} = f_0(\epsilon_{\vec{k}n}) + \delta f(\epsilon_{\vec{k}n}) \approx \Theta(\mu - \epsilon_{\vec{k}n}) + \delta f(\epsilon_{\vec{k}n})$, where $\Theta$ is a step function, μ the Fermi level $f_0$ is the Fermi distribution and $\delta f$ is a small perturbation which we approximate in our dynamics code as only depending on the energy (i.e. assuming momentum information is rapidly lost or irrelevant). We can write the change of emission relative to equilibrium (i.e. difference from the black-body spectrum) as:

$$\delta\Gamma_{\text{scatt,direct}}(\omega) = \int d\varepsilon \delta f(\varepsilon)\Gamma_{\text{direct}}(\varepsilon,\omega),$$

$$\Gamma_{\text{direct}}(\varepsilon,\omega) = \frac{4e^2\omega}{m_e^2 c^3}\text{Re}\{\sqrt{\epsilon_m(\omega)}\}\int_{BZ}\frac{g_s d\vec{k}}{(2\pi)^3}\sum_{n',n}\left(\delta(\varepsilon-\varepsilon_{\vec{k}n'})f_{\vec{k},n} - f_{\vec{k},n'}\delta(\varepsilon-\varepsilon_{\vec{k}n})\right)\delta(\varepsilon_{\vec{k}n'} - \varepsilon_{\vec{k}n} - \hbar\omega)\left|\vec{p}^{\vec{k}}_{\alpha,n,n'}\right|^2,$$

for direct transitions and

$$\delta\Gamma_{\text{scatt,phonon}}(\omega) = \int d\varepsilon \delta f(\varepsilon)\Gamma_{\text{phonon}}(\varepsilon,\omega),$$

$$\Gamma_{\text{phonon}}(\varepsilon,\omega) = \frac{4e^2\omega}{m_e^2 c^3}\text{Re}\{\sqrt{\epsilon_m(\omega)}\}\int_{BZ}\frac{g_s d\vec{k}'d\vec{k}}{(2\pi)^6}\sum_{n'n\alpha\pm}\left(\delta(\varepsilon-\varepsilon_{\vec{k}n'})f_{\vec{k},n} - f_{\vec{k},n'}\delta(\varepsilon-\varepsilon_{\vec{k}n})\right)\left(n_{\vec{k}'-\vec{k},\alpha} + \frac{1}{2}\mp\frac{1}{2}\right)\delta(\varepsilon_{\vec{k}'n'} - \varepsilon_{\vec{k}n} - \hbar\omega \mp \hbar\omega_{\vec{k}'-\vec{k},\alpha})\left|\sum_{n_1}\left(\frac{g^{\vec{k}'-\vec{k},\alpha}_{\vec{k}'n',\vec{k}n_1}\vec{p}^{\vec{k}}_{n_1 n}}{\varepsilon_{\vec{k}n_1} - \varepsilon_{\vec{k}n} - \hbar\omega + i\eta} + \frac{\vec{p}^{\vec{k}'}_{n'n_1}g^{\vec{k}'-\vec{k},\alpha}_{\vec{k}'n_1,\vec{k}n}}{\varepsilon_{\vec{k}'n_1} - \varepsilon_{\vec{k}n} \mp \hbar\omega_{\vec{k}'-\vec{k},\alpha} + i\eta}\right)\right|^2$$

for phonons. The total post-scattered luminescence can therefore be written as

$$\delta\Gamma_{\text{scatt}}(\omega) = \int d\varepsilon \delta f(\varepsilon)(\Gamma_{\text{direct}}(\varepsilon,\omega) + \Gamma_{\text{phonon}}(\varepsilon,\omega)).$$

We note that for both direct and phonon assisted transitions in $\left(\delta(\varepsilon-\varepsilon_{\vec{k}n'})f_{\vec{k},n} - f_{\vec{k},n'}\delta(\varepsilon-\varepsilon_{\vec{k}n})\right)$ the first term corresponds to unexcited electrons and the second to unexcited holes. Consequently, we can define a quantity just like the carrier distribution output which contains the carrier-resolved emission contribution. We present $\Gamma_{\text{direct}}(\varepsilon,\omega) + \Gamma_{\text{phonon}}(\varepsilon,\omega)$ in Figure S8. Once we evaluate this once, we can integrate against $\delta f$ for a specific perturbed case to evaluate its emission. $\delta f$ is discussed further in section 3.

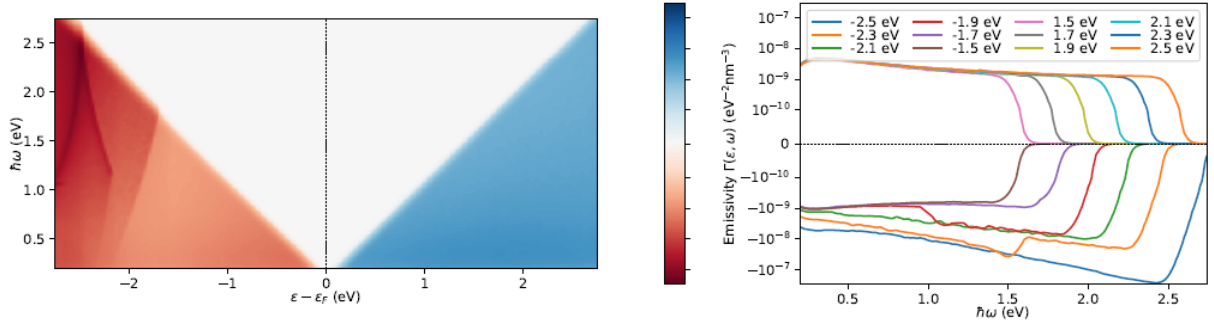

*Figure S8. Emissivity, $\Gamma_{\text{direct}}(\varepsilon, \omega) + \Gamma_{\text{phonon}}(\varepsilon, \omega)$, as a function of emission energy ($\hbar\omega$) and excited electron/hole energy ($\epsilon - \epsilon_F$). Plot on right presents lineslices of main plot at different electron/hole energies, as marked on legend.*

2. <u>Pre-scattered carriers</u>

The above approximation of working with $\delta f$ may be inaccurate for emission from excited carriers prior to any scattering. We can account for this by estimating the emission rate from the unscattered carriers prior to scattering by getting back to the same initial state. This would emit a photon with the same nominal frequency as the absorption (except for broadening). However, due to short hole lifetimes the broadening effect is significant. We can therefore collect contributions corresponding to unscattered carriers at energy $\varepsilon$ due to absorption and emission of photons both at frequency $\omega$.

Starting from the expression for $\text{Im}(\epsilon_m(\omega))$ above, the rate of change in electron and hole occupations due to absorption rate $a(\omega)$ of a photons of frequency $\omega$ per unit volume is (written with the $n$ and $n'$ contributions separately for simplicity using accumulation operators):

$$\dot{f}_{\vec{k},n} \mathrel{-}= \frac{a(\omega)}{\text{Im}(\epsilon_m(\omega))} \cdot \frac{4\pi^2 e^2}{m_e^2 \omega^2} \sum_{n'} (f_{\vec{k},n} - f_{\vec{k},n'}) \delta(\varepsilon_{\vec{k}n'} - \varepsilon_{\vec{k}n} - \hbar\omega) \left|\vec{p}^{\vec{k}}_{\alpha,n',n}\right|^2$$

$$\dot{f}_{\vec{k},n'} \mathrel{+}= \frac{a(\omega)}{\text{Im}(\epsilon_m(\omega))} \cdot \frac{4\pi^2 e^2}{m_e^2 \omega^2} \sum_{n'} (f_{\vec{k},n} - f_{\vec{k},n'}) \delta(\varepsilon_{\vec{k}n'} - \varepsilon_{\vec{k}n} - \hbar\omega) \left|\vec{p}^{\vec{k}}_{\alpha,n',n}\right|^2.$$

With carrier lifetimes $\tau_{\vec{k},n}$, this results in a steady-state change of occupation given by:

$$\delta\dot{f}_{\vec{k},n} \mathrel{-}= \tau_{\vec{k},n} \frac{a(\omega)}{\text{Im}(\epsilon_m(\omega))} \cdot \frac{4\pi^2 e^2}{m_e^2 \omega^2} \sum_{n'} (f_{\vec{k},n} - f_{\vec{k},n'}) \delta(\varepsilon_{\vec{k}n'} - \varepsilon_{\vec{k}n} - \hbar\omega) \left|\vec{p}^{\vec{k}}_{\alpha,n',n}\right|^2$$

$$\delta\dot{f}_{\vec{k},n'} \mathrel{+}= \tau_{\vec{k},n'} \frac{a(\omega)}{\text{Im}(\epsilon_m(\omega))} \cdot \frac{4\pi^2 e^2}{m_e^2 \omega^2} \sum_{n'} (f_{\vec{k},n} - f_{\vec{k},n'}) \delta(\varepsilon_{\vec{k}n'} - \varepsilon_{\vec{k}n} - \hbar\omega) \left|\vec{p}^{\vec{k}}_{\alpha,n',n}\right|^2.$$

The corresponding change in emission rate of photons per unit volume and unit $\omega'$ from the same channel (i.e. just reversing the above absorption) is:

$$\delta\Gamma_{\text{pre-scatt}}(\omega') = \frac{4e^2\omega'}{m_e^2 c^3} \text{Re}\{\sqrt{\epsilon_m(\omega')}\} \int_{BZ} \frac{g_s d\vec{k}}{(2\pi)^3} \sum_{n',n} (\delta f_{\vec{k},n'}(1-f_{\vec{k},n}) - f_{\vec{k},n'}\delta f_{\vec{k},n})\delta(\varepsilon_{\vec{k}n'} - \varepsilon_{\vec{k}n} - \hbar\omega')\left|\vec{p}^{\vec{k}}_{\alpha,n,n'}\right|^2$$

$$\delta\Gamma_{\text{pre-scatt}}(\omega') = \frac{a(\omega)}{\text{Im}(\epsilon_m(\omega))} \cdot \frac{4\pi^2 e^2}{m_e^2 \omega^2} \int_{BZ} \frac{g_s d\vec{k}}{(2\pi)^3} \sum_{n',n} (f_{\vec{k},n} - f_{\vec{k},n'})\delta(\varepsilon_{\vec{k}n'} - \varepsilon_{\vec{k}n} - \hbar\omega)\left|\vec{p}^{\vec{k}}_{\alpha,n',n}\right|^2$$

$$\times \frac{4e^2\omega'}{m_e^2 c^3} \text{Re}\{\sqrt{\epsilon_m(\omega')}\}(\tau_{\vec{k},n'}(1-f_{\vec{k},n}) + f_{\vec{k},n'}\tau_{\vec{k},n})\delta(\hbar\omega - \hbar\omega')\left|\vec{p}^{\vec{k}}_{\alpha,n',n}\right|^2$$

$$\delta\Gamma_{\text{pre-scatt}}(\omega') = \frac{1}{\frac{(\omega-\omega')^2}{\tau} + 1} \frac{a(\omega)}{\text{Im}(\epsilon_m(\omega))} \cdot \frac{4\pi^2 e^2}{m_e^2 \omega^2} \int_{BZ} \frac{g_s d\vec{k}}{(2\pi)^3} \sum_{n',n} (f_{\vec{k},n} - f_{\vec{k},n'})\delta(\varepsilon_{\vec{k}n'} - \varepsilon_{\vec{k}n} - \hbar\omega)\left|\vec{p}^{\vec{k}}_{\alpha,n',n}\right|^2$$

$$\times \frac{4e^2\omega'}{m_e^2 c^3} \text{Re}\{\sqrt{\epsilon_m(\omega')}\}(\tau_{\vec{k},n'}(1-f_{\vec{k},n}) + f_{\vec{k},n'}\tau_{\vec{k},n})\left|\vec{p}^{\vec{k}}_{\alpha,n',n}\right|^2$$

As expected, the self-emission has to be at the same frequency (up to broadening due to $\tau_{\vec{k},n}$). We can therefore calculate the total self-emission ($\int d\omega'$ of above) and add in the Lorentzian frequency dependence. Normalized per photon absorbed, the self-emission probability (dimensionless) is therefore:

$$\frac{\delta\Gamma_{\text{pre-scatt}}(\omega')}{a(\omega)} = \frac{1}{\text{Im}(\epsilon_m(\omega))} \frac{4\pi^2 e^2}{\hbar m_e^2 \omega^2} \int_{BZ} \frac{g_s d\vec{k}}{(2\pi)^3} \sum_{n',n} (f_{\vec{k},n} - f_{\vec{k},n'})\left|\vec{p}^{\vec{k}}_{\alpha,n',n}\right|^2$$

$$\times \frac{4e^2\omega'}{m_e^2 c^3} \text{Re}\{\sqrt{\epsilon_m(\omega')}\}(\tau_{\vec{k},n'}(1-f_{\vec{k},n}) + f_{\vec{k},n'}\tau_{\vec{k},n})\left|\vec{p}^{\vec{k}}_{\alpha,n',n}\right|^2 \frac{\gamma_{\vec{k}n'n}}{\pi\left[(\omega-\omega')^2 + \gamma_{\vec{k}n',n}^2\right]},$$

where the Lorentzian broadening is based on the inverse carrier lifetimes $\gamma_{\vec{k}n'n} = \frac{\tau_{\vec{k}n'}^{-1} + \tau_{\vec{k}n}^{-1}}{2}$ due to electron-electron and electron-phonon scattering (this emerges from the imaginary part of the self energy, $\text{Im}\Sigma_{\vec{k}n} = \frac{\hbar\tau_{\vec{k}n}^{-1}}{2}$, of each carrier in the energy-conserving $\delta$-function). Finally, we note that $PL_{internal} = \delta\Gamma_{\text{scatt}} + \delta\Gamma_{\text{pre-scatt}}$. $\delta f$ is discussed further in section 3.

We do not account for phonon-assisted transitions in the self-emission as only direct transitions would be sensitive to k-point smearing. We present $\delta\Gamma_{\text{pre-scatt}}(\omega)$ in Figure S9.

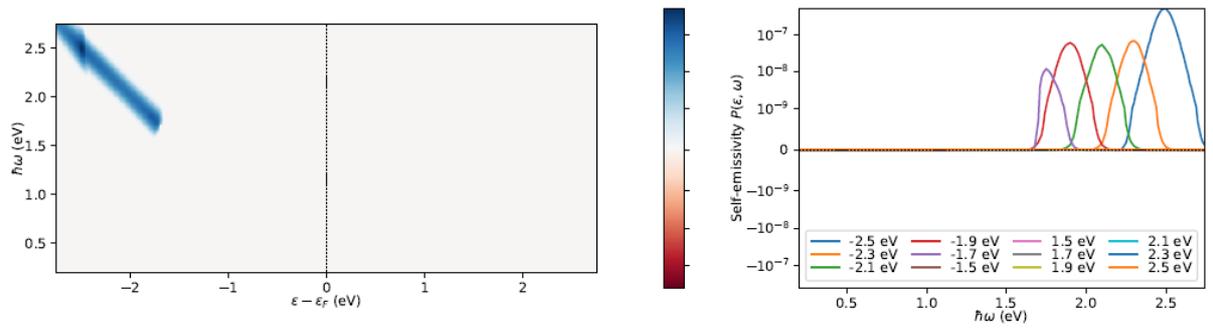

*Figure S9. Self-emissivity, $\delta\Gamma_{\text{pre-scatt}}(\omega)$, as a function of emission energy ($\hbar\omega$) and excited electron/hole energy ($\epsilon - \epsilon_F$). Plot on right presents lineslices of main plot.*

3. Calculating $\delta f(\varepsilon)$

As shown in the main text (Figure 1g) the photoluminescence is a first order process. To calculate $\delta f(\varepsilon)$ we take equation 2 of [11], the Boltzmann equation with electron-electron and electron-phonon scattering parameterized entirely from DFT calculations (noting electron-electron scattering includes what is sometimes referred to as Auger effects [12]) and add in a continuous wave heating term:

$$\frac{d}{dt}f(\varepsilon, t) = \Gamma_{e-e}[f](\varepsilon) + \Gamma_{e-ph}[f, T_l](\varepsilon) + \frac{p_{abs}P(\omega, \varepsilon)}{g(\varepsilon)},$$

where

$$\Gamma_{e-e}[f](\varepsilon) = \frac{2D_e}{\hbar} \int \frac{d\varepsilon_1 d\varepsilon_2 d\varepsilon_3 \big(g(\varepsilon_1)g(\varepsilon_2)g(\varepsilon_3)\big)}{g^3(\varepsilon_F)}$$
$$\times \delta(\varepsilon + \varepsilon_1 - \varepsilon_2 - \varepsilon_3)\{f(\varepsilon_2)f(\varepsilon_3)\big(1 - f(\varepsilon)\big)\big(1 - f(\varepsilon_1)\big)$$
$$- f(\varepsilon)f(\varepsilon_1)\big(1 - f(\varepsilon_2)\big)\big(1 - f(\varepsilon_3)\big)\},$$

with $D_e$ a constant of proportionality extracted from ab-initio calculations of electron lifetimes [6],

$$\Gamma_{\text{e-ph}}[f, T_l](\varepsilon) = \frac{1}{g(\varepsilon)}\frac{\partial}{\partial\varepsilon}\left[H(\varepsilon)\left(f(\varepsilon)\big(1 - f(\varepsilon)\big) + k_B T_l \frac{\partial f}{\partial \varepsilon}\right)\right],$$

where $H(\varepsilon)$ is an energy-resolved electron-phonon coupling strength calculated from ab-initio electron-phonon matrix elements [6], $p_{abs}$ is the absorbed power density, $P(\omega_{in}, \varepsilon)$ is carrier distribution excited by a single photon of energy $\hbar\omega_{in}$ (as defined in Brown et al. [6]), $g(\varepsilon)$ is the electronic density of states (to convert from carrier number to occupation change).

We set $\frac{df}{dt} = 0$ and perform an expansion in $f(\varepsilon)$ to linearise the right hand side. The solution to the linearised version of this equation gives the quantity $\frac{\delta f(\varepsilon)}{p_{abs}}$, which as a last step we multiply by $\hbar\omega_{in}$ to give a quantity per absorbed photon in subsequent analyses. We note this approach is equivalent to Sivan et al.'s calculation but here interband transition elements are also included in our calculations [13]. We present $\frac{\delta f(\varepsilon)}{p_{abs}}$ in Figure S10.

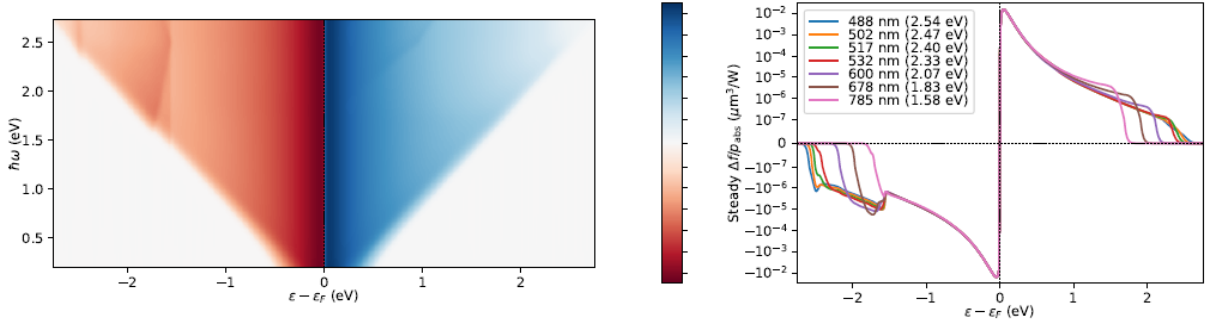

*Figure S10. Steady state electron/hole population per absorbed photon as a function of laser excitation energy ($\hbar\omega$) and excited electron/hole energy ($\varepsilon - \varepsilon_F$).*

4. <u>Confirming black body emission theory</u>

Here we demonstrate that our equations reduce to black body emission theory for spontaneous emission in the case of no excitation (i.e. confirm thermodynamic equilibrium). In equation S7 for emission of light from a point in the metal, we note that $\epsilon_{\vec{k}n'} = \epsilon_{\vec{k}n} + \hbar\omega$ and that, at equilibrium

$$f_{\vec{k}n'}(1 - f_{\vec{k}n}) = \frac{f_{\vec{k}n} - f_{\vec{k}n'}}{e^{\hbar\omega\beta} - 1}$$

where $\beta = \frac{1}{k_B T}$, $k_B$ is the Boltzmann constant and $T$ temperature. Applying this transformation we can write the emission per unit volume, per unit energy, per unit time in the material at equilibrium as

$$\Gamma_{\text{equib}}(E) = \frac{\text{Re}\{\sqrt{\epsilon_m(\omega)}\}\omega}{\hbar\pi^2 c^3} \times \frac{1}{e^{\hbar\omega\beta}-1} \times \text{Im}(\epsilon(\omega_{out})) = \frac{8\pi\alpha \text{Re}\{\sqrt{\epsilon_m(\omega)}\}^2 E^2}{h^3 c^2(e^{E\beta}-1)}.$$

The emission per unit volume at equilibrium is also given by the Shockley-Van-Roosebroek formula as $4\pi\alpha(E)\text{Re}\{\sqrt{\epsilon_m(\omega)}\}^2 \phi_{bb}(E)$ (the definition of internal luminescence in semiconductors), which is identical to what we obtain above [14]. Here $\alpha(E)$ is the absorption coefficient of the gold (per unit length) and $\phi_{bb}$ is the black body emission flux per unit energy, per unit area, per unit solid angle.

5. Simulation results explored in more detail

There is a significant difference between $\text{Re}\{\sqrt{\epsilon_m(\omega)}\}$ from experiment, $\text{Re}\{\sqrt{\epsilon_m(\omega)}\}_{exp}$, and theory, $\text{Re}\{\sqrt{\epsilon_m(\omega)}\}_{th}$, especially at wavelengths longer than 600 nm (see Figure S11a). Therefore, when multiplying our DFT calculations by $\text{Re}\{\sqrt{\epsilon_m(\omega)}\}$ we used the experimental value for the plot in the main text. However, as noted in the previous section, for black-body equilibrium equation S7 reduces to $4\pi\alpha(E)\text{Re}\{\sqrt{\epsilon_m(\omega)}\}^2 \phi_{bb}(E)$. Therefore, it can be argued that there is still a discrepancy from the second $\text{Re}\{\sqrt{\epsilon_m(\omega)}\}$ factor. In Figure S11b we present the equivalent of Figure 3d in the main text but multiplying the luminescence by $\frac{\text{Re}\{\sqrt{\epsilon_m(\omega)}\}_{exp}}{\text{Re}\{\sqrt{\epsilon_m(\omega)}\}_{th}}$. It can be seen that there is better experiment-theory agreement, especially at longer wavelengths. Therefore, a possible cause of discrepancies at longer wavelengths due to the difference between the experimental and calculated values of $\text{Re}\{\sqrt{\epsilon_m(\omega)}\}$.

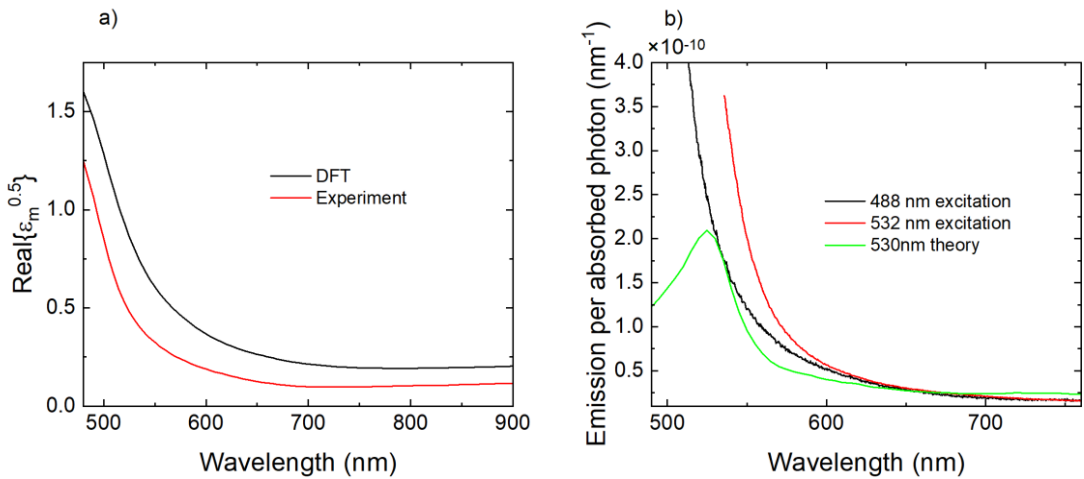

*Figure S11. a) DFT and experimental values of $n'$ as a function of wavelength. b) DFT prediction when multiplied by a factor of $\frac{\text{Re}\{\sqrt{\epsilon_m(\omega)}\}_{exp}}{\text{Re}\{\sqrt{\epsilon_m(\omega)}\}_{th}}$, alongside experimental results presented in Figure 3c.*

In Figure S12a we present the simulated internal luminescence for different excitation wavelengths. For excitation at 490 nm a much stronger luminescence signal is predicted at shorter wavelengths than is observed experimentally (see Figure S10). This is due to the simulation including a higher energy d-band that is not present in the experiment. We note this discrepancy is well within the tolerance for a DFT calculation. The luminescence signal can be broken down into two parts – pre- and post-scattered luminescence, as is presented in Figure S12b for 530 nm excitation.

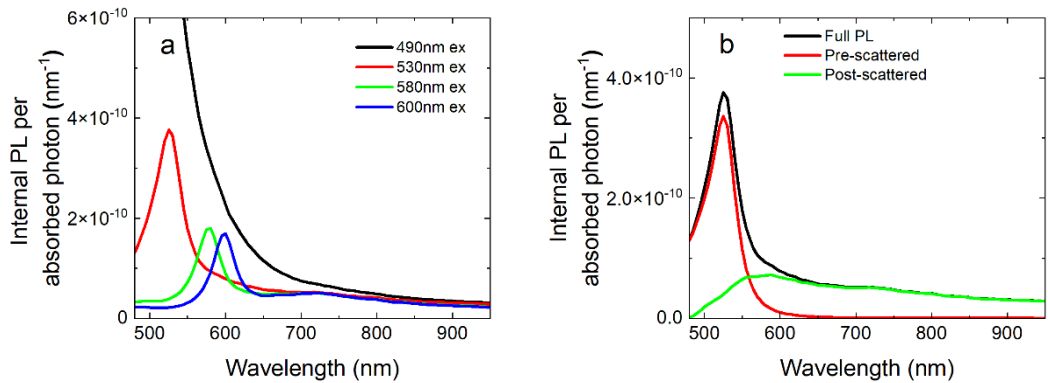

*Figure S12. a) Predicted internal luminescence for different excitation wavelengths. b) The breakdown of internal luminescence into pre-scattered and post-scattered luminescence for 530 nm excitation.*

In Figure S13 we present the total DFT simulated PL in energy space. We note that to generate this figure we used $\text{Re}\{\sqrt{\epsilon_m(\omega)}\}_{th}$ instead of $\text{Re}\{\sqrt{\epsilon_m(\omega)}\}_{exp}$ as it is parametrized much further into the infra-red region.

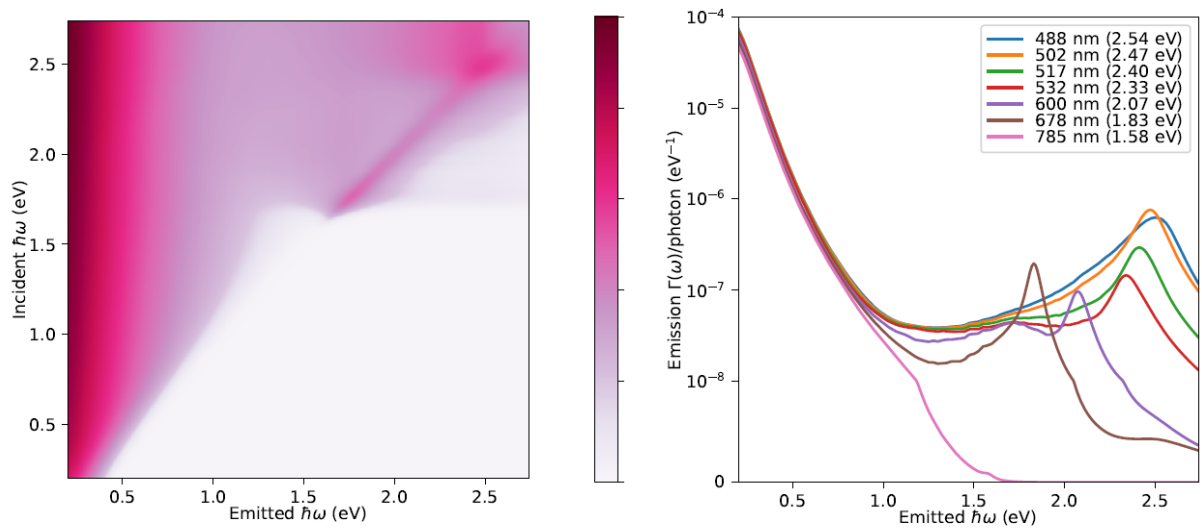

*Figure S13. Steady state photoluminescence as a laser (incident) excitation energy (ℏω) and emitted luminescence energy. Plot on right hand side presents lineslices for specific excitation wavelengths.*

**Supplemental Note 9 – Recording the same signal when exciting and observing the signal from the glass side or air side**

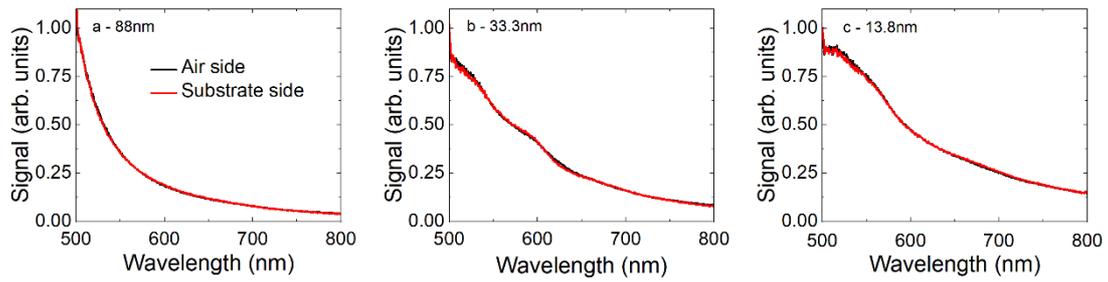

*Figure S14. Photoluminescence signal when sample is excited and signal is observed from the air side (black curve), or through the substrate (red curve), with 488 nm excitation, and for different sample thicknesses as indicated on the inset. Incident intensity was 0.017 mWµm$^{-2}$ when exciting through substrate and 0.079 mWµm$^{-2}$ when exciting from air side.*

**Supplemental Note 10 – Further discussion of thin flake photoluminescence**

The model presented in the main text originates as follows: the wavevector in the direction perpendicular to the interface becomes discretised with $\frac{2\pi}{d}$ spacing in k-space. This gives a corresponding energy of $\hbar v_f n \times \frac{2\pi}{d} = \frac{hnv_f}{2d}$, where $v_f$ is the fermi velocity and $n$ is in integer spacing.

We note that there are other ways to fit the data presented in Figure 4c. We present a different fit via an experimentally motivated approach in Figure S15. Specifically, all energy shifts via a single equation: $Energy\ shift = ntK$, where $n$ is an integer (in the plot we observe $n =$1, 2, 4 and 6) and $K$ a fitting parameter which we record as (6.8±0.1) meV nm$^{-1}$. In the Figure S15 inset we plot energy shifts divided by $n$: it can be see that all resonance peaks lie on one straight line.

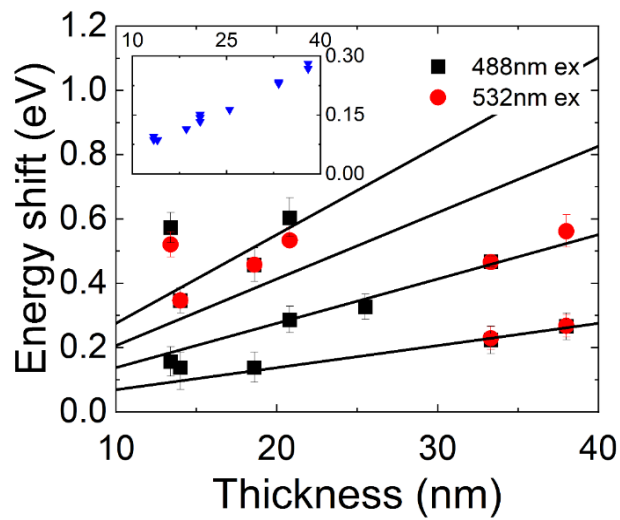

*Figure S15. Alternative linear fitting of the data presented in Figure 4c (see text for details).*

**Supplemental Note 11 – DFT calculations of the thickness-dependent electronic structure**

To further study band-structure effects on the observed photoluminescence, we performed first-principles calculations based on density-functional theory (DFT) on films with different thicknesses up to 40 (111) atomic planes. Calculation details are given in a recent study [15]. After calculating the Kohn-Sham electronic structure and eigenvalues using the Quantum Espresso package [16], we obtain transition dipole matrix elements using the YAMBO program [17] at the 21 nearest k-points around the M point of the Brillouin zone. Subsequently, we calculate the square of dipole matrix elements, $d^2_{\varepsilon,\varepsilon'} = |\langle\varepsilon|x|\varepsilon'\rangle|^2 + |\langle\varepsilon|y|\varepsilon'\rangle|^2$, as a function of initial and final energy states and averaged over k-points in the vicinity of the M point using the following equation

$$|\langle\varepsilon|x|\varepsilon'\rangle|^2 = \frac{\sum_{n,m,\boldsymbol{k}} \delta(\varepsilon' - \varepsilon_{m\boldsymbol{k}})\delta(\varepsilon - \varepsilon_{n\boldsymbol{k}})|\langle n\boldsymbol{k}|x|m\boldsymbol{k}\rangle|^2}{\sum_{n,m,\boldsymbol{k}} \delta(\varepsilon_i - \varepsilon_{m\boldsymbol{k}})\delta(\varepsilon_f - \varepsilon_{n\boldsymbol{k}})},$$

where delta functions are replaced by a normalized Gaussian of width $\sigma = 50$ meV. We plot $d^2_{\varepsilon,\varepsilon'}$ for different layer thicknesses in Figure S16.

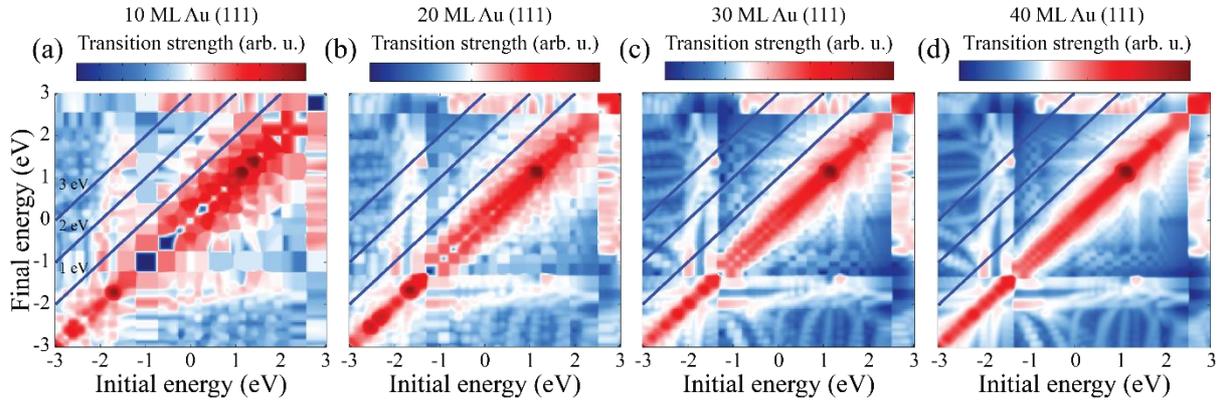

*Figure S16. Averaged dipole matrix elements of Au (111) films consisting of a) 10 atomic layers, b) 20 atomic layers, c) 30 atomic layers, and d) 40 atomic layers. Blue lines are drawn to indicate transition energies between initial and final states corresponding to 1, 2 and 3 eV. Blue and red colours in the colour bar illustrate the minimum and maximum transition strengths, respectively.*

**Supplemental Note 12 – Results under excitation with a laser of 785 nm wavelength**

Figure 1a in the main text shows that it was possible to measure luminescence from gold flakes following 785 nm excitation. We presented the signal per absorbed photon, where the absorption was based on the simulated absorption value (due to a larger error in the experimental value). We also note that for thinner flakes sample absorption is much stronger (and thus carries less error) in this wavelength region.

Here we discuss this measurement and its interpretation in detail. We begin by noting that this signal was extremely challenging to measure – gold's absorption coefficient at this wavelength is less than 5 % for all thicknesses, and the luminescence efficiency is less than $10^{-10}$ at all wavelengths. This meant that even small noise from the instrument competed with the observed signals. We present recorded luminescence signals in Figure S17a for four flake thicknesses. The luminescence increases for thinner flakes, in agreement with our model. We found that a small portion of the observed signal originated from noise from within the microscope rather than the sample, with peaks at 870 nm and 900 nm (marked 'instrument' on the figure). Despite exploring different microscope configurations, we were unable to fully remove these signals. These peaks had a different spatial spread to the gold luminescence and were also observed when measuring a silver mirror, confirming they did not originate from the gold. To first approximation we can describe the signal as $Signal = Signal_{gold} + Signal_{instrument}$, with the second term approximately constant as flake thickness changes. To deconvolute these two components we subtracted the signal from two flakes of different thicknesses i.e. $Signal(33.3nm) - Signal(88nm) = Signal_{gold}(33.3nm) - Signal_{gold}(88nm)$. From this point we proceeded with the same analysis as presented in the main text.

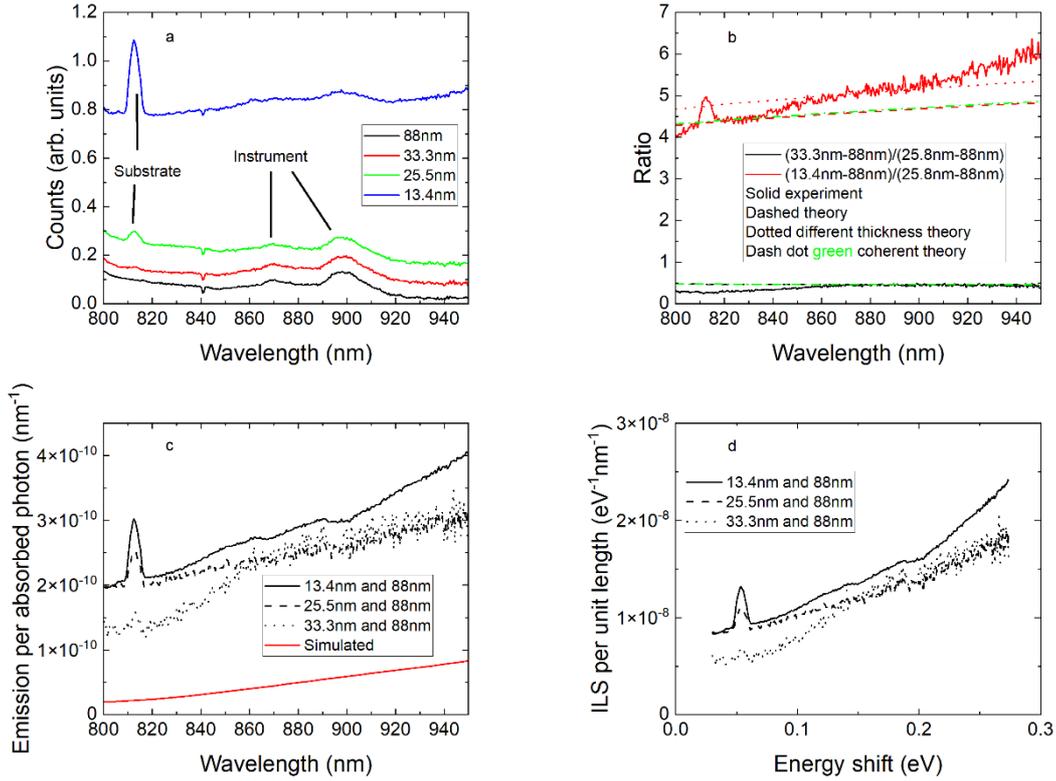

*Figure S17. a) Signal counts per incident power following 785 nm excitation at 1.87 mWμm$^{-2}$. Substrate raman peaks and additional signals from the instrument (see text) are marked. b) Ratios between signals presented in a), as described in the legend, alongside dashed lines based on our model prediction, and dotted lines assuming the thinnest flake is 1 nm thinner and other flake thicknesses unchanged. The theory for a fully coherent process is overlaid in green. c) the internal photoluminescence per absorbed photon, based on three different flake combinations (see figure). The simulated signal based on photoluminescence is in red. d) the strength of Inelastic Light Scattering (ILS) per unit length, as extracted from the three different flake combinations.*

We present the ratio of recorded signals in Figure S17b, specifically of the type $\frac{Signal(33nm) - Signal(88nm)}{Signal(25.8nm) - Signal(88nm)}$. Overlaid on this plot in dashed lines is the equivalent to equation 2 in the main text for the different thicknesses. Here we normalise the signals by that from 25.8 nm as this flake gives relatively strong signal and has less uncertainty in thickness than the 13.4 nm flake. However, even a small difference in the thickness can result in a significantly different predicted ratio: overlaid on this plot is a ratios setting the thinnest flake thickness to 12.5 nm (i.e., approximately 1 nm difference). This reveals that, within the experimental error of measured thicknesses (±2 nm), we find experiment and theory are in relatively good agreement. Importantly, unlike for the case of exciting at 488 nm or 532 nm, our theory agrees well with experiment for all thicknesses and we see no additional resonance features at longer wavelengths for thin flakes. We also overlay the predictions from theory assuming a fully coherent process – it can be seen that the results are almost identical. Figure S17b also demonstrates that our approach is relatively successful in removing the instrument contribution to the signal.

We used the same approach as in the main text to calculate the internal PL per absorbed photon, which we present in Figure S17c for three different flake combinations. In all cases we obtain a very similar shape to the internal PL. Overlaid on this plot in red is the internal PL predicted from our DFT-parameterized simulations. The simulated PL is approximately a factor of 5 weaker than the experimental signal.

We further confirm that the internal PL should be on the order of $10^{-11}$ without reference to DFT via a simple order of magnitude calculation. To first order, gold's absorption coefficient is the same strength at 488 nm and 785 nm, implying that the total transition rate is the same for the two cases. For 488 nm excitation, photon absorption only populates a small region in energy of the d-band. In contrast, when exciting at 785 nm all energies from 1.58 eV below the fermi level (785 nm in energy) to the fermi level are approximately equally populated (as this proceeds via phonon-assisted transitions). Finally, the scattering rate is approximately an order of magnitude lower for sp-band holes, meaning they have a 'lifetime' in their current state approximately 10 times longer. Assuming equal optical transition rates at between all energies, we can estimate the probability of luminescence within 0.1 eV of the excitation energy as being $2 \times 10 \times \left(\frac{0.1}{2 \times 1.58}\right)^2$ less probable for 785 nm excitation than 488 nm excitation. Here the factor of 2 comes from both electrons and holes contributing to the photoluminescence equally for 785 nm excitation, the factor of 10 from the lifetime and the squared factor from considering: i) the proportion of the transition strength for absorption which populates high energy holes; ii) the proportion of the emission transition strength which we can apportion to these high energy holes. This 'back of the envelope' shows, without DFT calculations, that we expect a photoluminescence strength close to the excitation wavelength to be at least an order of magnitude weaker following 785 nm excitation than for 488 nm and 532 nm excitation (as predicted in our DFT simulations on the plot).

Our data shows that the luminescence signal observed following 785 nm excitation is a factor of 5 larger than is reasonable if we assume its origin is luminescence. The only other process which could cause this signal is inelastic light scattering, ILS. Therefore, we suggest that exciting in the intraband regime primarily results in ILS. The equivalent to the model presented in the main text for ILS is

$$ILS_{\text{external}}(\lambda_{\text{out}}) = ILS_{\text{internal}}(\lambda_{\text{em}}, \lambda_{\text{out}}) \left(\frac{I_{\text{in}}(\lambda_{\text{in}})}{\hbar\omega_0}\right) \int \frac{f_{\text{abs}}(z, \lambda_{\text{in}}) f_{\text{emit}}(z, \lambda_{\text{out}})}{\alpha_0} dz$$

where $\alpha_0$ is the absorption coefficient at the incident laser energy, and $ILS_{\text{internal}}$ is the probability of internal inelastic light scattering both per unit wavelength and per unit length within gold. The main change compared to the main text is to state inelastic light scattering is per laser intensity locally, rather than per absorbed photon locally. It is more typical to present ILS effects as an energy shift, so in Figure S17d we present $ILS_{\text{internal}}$ as a function of energy shift for the three different flake combinations considered. We note this factor includes both ILS and PL contributions.